# An Analysis of European Data and AI Regulations for Automotive Organizations

Charlotte A. Shahlaei, Halmstad University and the University of Notre Dame

Nicholas Berente, University of Notre Dame

## Abstract

This report summarizes the European Union's series of data and AI regulations and analyzes them for managers in automotive vehicle manufacturing organizations. In particular, we highlight the relevant ideas of the regulations, including how they find their roots in earlier legislation, how they contradict and complement each other, as well as the business opportunities that these regulations offer.

The structure of the report is as follows. First, we address the GDPR as the cornerstone against which the requirements of other regulations are weighed and legislated. Second, we explain the EU Data Act since it directly addresses Internet of Things (IoT) for businesses in the private sector and imposes strict requirements on large data generators such as vehicle manufacturers. For manufacturers, compliance with the EU Data Act is a prerequisite for the subsequent legislation, in particular the EU AI Act. Third, we explain the Data Governance Act, Digital Services Act, Digital Markets Act, and EU AI Act in chronological order. Overall, we characterize European Union data regulations as a wave set, rooted in historical precedent, with important implications for the automotive industry.

# 1. Executive Summary

In the years after the European Union (EU) General Data Protection Regulation (*GDPR*), multiple EU-level legislations have emerged to expedite and regulate the digital transformation of the European market. These legislations include the EU Data Governance Act, Digital Services Act, Digital Markets Act, Data Act, and AI Act.

|  | Passed | Enacted |
|---|---|---|
| ● **GDPR** | April 2016 | May 2018 |
| ● **Data Governance Act** | June 2022 | September 2023 |
| ● **Digital Services Act** | October 2022 | February 2024 |
| ● **Digital Markets Act** | November 2022 | March 2024 |
| ● **Data Act** | November 2023 | September 2025 |
| ● **AI Act** | June 2024 | June 2026 |

It is clear that European automotive vehicle manufacturing organizations (vehicle manufacturers) are not facing a single legislation as in the days of only GDPR, but a wave set of tech-related regulations. This study describes how major data generators will benefit from a careful understanding of the interaction of various legislative instruments and anticipating the future challenges and opportunities. Vehicle manufacturers are a critical category of organization - they are the archetypal industrial manufacturer, and at the same time incredibly important data generators. They cover two of the fields designated by the EU as strategic in creating a single digital market: mobility and manufacturing. This analysis is based on the concerns and anticipations of vehicle manufacturing experts and offers insights into the interaction between different legislative instruments as a wave set rather than focusing on the details of a single one.

Takeaways from this work include:

- We highlight and summarize the key elements of each EU data-oriented regulation for vehicle manufacturers.

- We show why vehicle manufacturers need to go beyond reactive compliance approaches to EU data-oriented regulations and find ways to position themselves to thrive in this era of wave upon wave of data regulations.

- We point out how requirements from regulations are often synergistic, and foreshadowed before they become requirements. For example:

    - GDPR's requirement for data access and data portability was foreshadowed by the Directive 95/46/EC.

    - The Data Act's conditions for protection of IoT databases were foreshadowed by the Directive 96/9/EC (Database Directive).

    - The AI Act's requirements for risk assessment of AI systems that are safety components of a product was foreshadowed by EU Harmonization Legislation in 2019 in general, and by Type Approval Regulation and General Safety standards in particular for vehicle systems.

- We identify opportunities for vehicle manufacturers to generate value in innovative ways consistent with EU data-regulations. For example:

    - By benefiting from the legislative instruments in the Data Act, vehicle manufacturers can become designated third parties with whom the users choose to share their IoT data generated through the use of various IoT-related products and services such as phones,

music apps, navigation systems. This opportunity allows vehicle manufacturers to access data from large technology (i.e. "big tech") companies.

- By benefiting from the legislative instruments in the Digital Markets Act, vehicle manufacturers could potentially become less dependent on single core service providers. This loosened dependence on the applications and services of big tech allows vehicle manufacturers to gain more negotiating power in forming contracts with big tech companies that are usually core service providers to vehicle manufacturers.

- By benefiting from the legislative instruments in the Data Governance Act, vehicle manufacturers can access large, privileged datasets (e.g., biometrics) held by the public sector bodies for research purposes such as debiasing AI training datasets.

- We point out how data-oriented regulations may interact and contradict each other or create tension points that require vehicle manufacturers to take additional measures. For example:

  - The legal bases for processing and sharing personal data regulated under the GDPR are not the same as those for processing and sharing IoT data regulated under the Data Act. Vehicle manufacturers must ensure that different categories of data (which are sometimes mixed) are connected to the right bases for processing and sharing.

  - Although some regulations such as the Digital Services Act do not directly apply to vehicle manufacturers, they can create tensions due the vehicle manufacturers' association with organizations that are addressed directly by these regulations. For instance, the Digital Services Act mandates a set of strict requirements on big tech companies regarding illegal content and activities. Some big tech companies are providers of telematics and navigation systems, applications and services to vehicle manufacturers. Per GDPR requirements, vehicle manufacturers are required to provide users with transparency on how the design of vehicle's systems and components can affect the collection, processing and protection of users' data by default. Thus, vehicle manufacturers must ensure that they are informed on the way the requirements in regulations such as the Digital Services Act can potentially interact with GDPR requirements.

  - The GDPR requires vehicle manufacturers to provide data subjects with information such as the time, duration and purposes of processing their data. This specific GDPR requirement, however, was composed having personal data categories in mind. Personal data categories are more static categories of data, unlike IoT data flows. At the same time, to stay compliant with requirements of regulations on AI systems and automatic functions in vehicles, manufacturers need to process live IoT data flows on-board vehicles. Compliance with the GDPR requirements, then, requires vehicle manufacturers to provide vehicle users with information about the processing of their IoT data without having to constantly intrude the driver or user with information or consent requests for processing a continuous flow of IoT data generated throughout the vehicle travel time.

It is important to note that this document is not intended to give legal advice nor take the place of legal counsel. We are not lawyers; we are business researchers that focus on digital innovation. The goal of this work is for business professionals to think through the important opportunities and constraints that present themselves through the wave upon wave of data-oriented regulations that are the new normal for the new generation of vehicle manufacturing. It is no longer enough for business leaders in vehicle manufacturing to delegate regulatory issues to their legal staff. Of course, legal staff should be key stakeholders involved with organizational decisions, but it is critical that executives, managers, and engineers also understand the legal and technical implications of the EU regulatory wave set in their product innovation and strategic planning decisions.

## 2. Overview

Vehicle manufacturers generate considerable amounts of data. Yet, their business models have formed around manufacturing physical products, and data has only recently become a strategic focus for these organizations. Historically, they typically used data for optimizing operations, processes, and service development. Most vehicle manufacturers do not leverage data like large technology companies. But this is changing. Vehicle manufacturers generate and have access to significant datasets and are subject to the strictest requirements by the recent EU regulations both directly (e.g., required to conduct meticulous risk assessment of AI systems), and indirectly (e.g., their business relationship with tech giants turns them into high-risk data controllers and AI deployers). Further, vehicle manufacturers are increasingly seeing the strategic value of data. Vehicle manufacturers need to go beyond the reactive compliance methods and are adopting comprehensive governance approaches that enable them to generate value from their data as they seek to understand and leverage the synergies among different legislations.

The European Union (EU) is setting many regulatory standards for digital data and artificial intelligence (AI). Many organizations around the globe need to address these regulations in some form or another. The EU's strategy is to create a single market for data and establish open and common data spaces for a number of strategic fields such as health, agriculture, mobility, manufacturing, energy, finance, and public administration. The strategy is focused on making data available for innovation and sustainability in these sectors, while keeping organizations that generate data in check. In this effort, the EU has devised a series of major legislative instruments. For instance, in the years after the General Data Protection Regulation (*GDPR*) mandate, specifically in the last two years, multiple EU legislations have emerged to expedite and regulate the digital transformation of the European market. These legislations include the *Data Governance Act* (passed June 2022 - fully applicable September 2023), *Digital Services Act* (passed October 2022 - fully applicable February 2024), *Digital Markets Act* (passed November 2022 - fully applicable March 2023), *Data Act* (passed November 2023 - fully applicable September 2025), and *AI Act* (passed June 2024 fully applicable June 2026; see Figure 1).

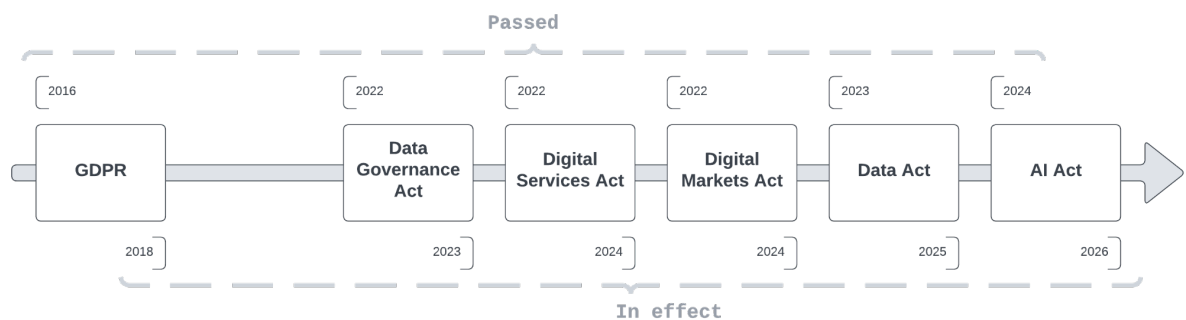

*Figure 1. Progression of Europe's Recent and Leading Regulations on Data and Information Technology*

Organizations that interact with Europe face multiple mandates that often require them to change their technical, governance, and organizational approaches to data and AI. The simultaneous application of different legislative instruments across various sectors reveals potential synergies, but also potential conflicts.

These Union-level legislation have three major instrumental levers.:

1. They create the *legal* bases and mechanisms to allow data to flow freely among various players across sectors.

2. They specify the *technical* infrastructure to ensure that data sharing and processing happens in a safe and secure way. This provides the technical assistance for actors to achieve compliance with mandatory requirements or have enough encouragement to take part in voluntary data sharing.

3. They control the market power of large enterprises with sizable datasets and user networks to create *economic* incentives for less powerful actors in the market. These regulations have a tiered way with both identifying the scope of control and risks that various actors' operations imply when collecting and processing data, and with subjecting them to corresponding responsibility.

This primer describes each legislation by first providing an overview, and then explaining how the legislation may affect automotive organizations. This analysis is based on the concerns and anticipations of vehicle manufacturing experts and concludes with insight into the interaction among the different legislations rather than focusing on the details of any single legal instrument. It is important to reiterate that this is not a legal document, nor do we intend to give legal advice. We do not claim that our account is comprehensive, and different legislation can and may be interpreted by others differently than we characterize it. The goal is not to provide an exhaustive account of the regulations, but rather, a guide and analysis of some of the main issues and concerns to vehicle manufacturers. This is a report intended for managers, and of course legal experts should be consulted before making organizational decisions based on the ideas herein.

# 3. General Data Protection Regulation (GDPR)

Three regulations constitute Europe's data protection base: the *General Data Protection Regulation* or the *GDPR* (personal data privacy in the context of the EU's private sector)[1], *Regulation (EU)2018/1725* (personal data privacy in the context of the EU's public institutions and agencies)[2], and *Directive (EU) 2016/680* (personal data privacy in the context of law enforcement)[3]. Except for a few exceptions, both the regulation and the directive closely match the requirements of the GDPR. Here, we focus on the GDPR as it pertains to the private sector (and hence the automotive manufacturers).

## 3.1. The GDPR and Directive 95/46/EC

Personal data privacy has always been an important issue. Before the GDPR, previous legislative instruments such as Directive 95/46/EC[4] had also aimed at protecting personal data for decades. However, in 2016, the European Parliament and Commission proposed that the current speed of technological change called for a review of the existing categories, definitions, and requirements related to the protection of personal data. For instance, the GDPR counts several categories of data that were not covered by Directive 95/46/EC. Further, Directive 95/46/EC only focused on the regulation of *data controllers* – the people and organizations that determine the purposes and means of processing personal data. These did not include *data processors* – those who process personal data on behalf of the controller. Under the GDPR, data processors must also comply with similar requirements to those that are imposed on data controllers, albeit in a slightly more limited way.

It is worth mentioning that, as a directive (not a regulation), Directive 95/46/EC also set the goals that must be achieved in protecting personal data, but it allowed EU Member States the freedom to decide how to translate directives into national laws. This freedom has, over the course of several decades, created fragmentation in how EU Member States protect personal data. As a result, complying with various personal data protection requirements has turned into a lengthy and costly undertaking for companies operating across the EU. For instance, when it comes to breach notification obligations, Directive 95/46/EC did not contain a specific obligation for organizations to notify data breaches. Therefore, various Member States treated the matter differently with, for instance, Germany and Italy, requiring organizations to notify certain categories of breaches, and others like the Netherlands requiring a notification of all personal data breaches.[5]

## 3.2. Summary of the GDPR

On a general level, the GDPR aims to remove fragmentation and unify strategies for protecting personal data across various EU Member States. It legislates that organizations adopt privacy protection obligations to safeguard people's fundamental rights in the face of new technologies. Overall, the goals involve giving people more control over their personal data. For example, the GDPR provides rights to access data in a machine-readable way, rights to data portability, rights to be notified in case of a data breach, the right to be forgotten or data erasure upon request (when there is not legitimate reason to keep their data), and the right not to be subject to a decision based solely on automated means.

---

[1] Regulation (EU) 2016/679 of the European Parliament and of the Council of 27 April 2016 on the protection of natural persons with regard to the processing of personal data and on the free movement of such data, and repealing Directive 95/46/EC (General Data Protection Regulation).

[2] Regulation (EU) 2018/1725 of the European Parliament and of the Council of 23 October 2018 on the protection of natural persons with regard to the processing of personal data by the Union institutions, bodies, offices and agencies and on the free movement of such data, and repealing Regulation (EC) No 45/2001 and Decision No 1247/2002/EC.

[3] Directive (EU) 2016/680 of the European Parliament and of the Council of 27 April 2016 on the protection of natural persons with regard to the processing of personal data by competent authorities for the purposes of the prevention, investigation, detection or prosecution of criminal offenses or the execution of criminal penalties, and on the free movement of such data.

[4] Directive 95/46/EC of the European Parliament and of the Council of 24 October 1995 on the protection of individuals with regard to the processing of personal data and on the free movement of such data.

[5] Article 29, Data Protection Working Party.

GDPR defines personal data broadly. The following categories of information are considered to be personal data and complement the categories of personal data already existing in Directive 95/46/EC:

- photos and audio/visual formats
- financial transactions
- posts on social networking websites
- device identifiers (computer's MAC/IP address, mobile phones IMEI number)
- location data
- user login credentials
- browsing history
- genetic information

The introduced changes can be outlined around several main principles devised to enhance the protection of personal data: proactive accountability, a risk-based approach, and data protection by design.

**3.2.1. Proactive Accountability**

Fines for GDPR violations can be quite steep. However, the regulation states that the fines must be proportionate, taking the violation's severity and scope into account. One of the ways to judge such proportionality is by taking into consideration the efforts of an organization to be compliant and to actively demonstrate compliance. GDPR encourages several strategies to help demonstrate an organizations' compliance efforts. These strategies include, for instance, the appointment of data protection officers (DPOs), abiding by codes of conduct, and acquiring seals and certifications of compliance. The GDPR also recognizes that in order for such strategies to be worthwhile, they need to provide practical advantages. Some examples include the following measures that are not mandatory but can help organizations with procedural, technical, and reputational issues related to data protection.

- **The appointment of DPOs:** is mandatory only in case a data controller/processor conducts regular and systematic monitoring of data subjects or sensitive data on a large scale (e.g., vehicle manufacturers). A DPO has special qualifications and special protections to act independently and provide the organization with the primary contact point for data protection issues. At the same time, since the DPO has special qualifications and expertise, they can provide the business with consultation and review of all the data protection issues and support the procedural planning processes.

- **Adhering with codes of conduct:** is legally binding and at the same time beneficial for businesses because it provides a means for certain industry sectors, or groups of organizations, to create context-specific rules regarding the processing of personal data. It could, for instance, make impact assessment easier since one can benchmark their own approach against the approach of other actors within the same industry to detect best practices. Once the adherence with a code of conduct is formally approved by a data protection authority (DPA), it can also provide the basis for cross-border data transfers if the non-EEA organization can show adherence to code of conduct. Additionally, it can affect the fines imposed upon the adherent controller or processor since they have proven substantial efforts for compliance through adhering with code of conduct.

- **Seals and certifications:** have reputational boosts and can be granted (in the form of a visual symbol such as badges/emblems that can be displayed on published documents or websites) to confirm that organizations satisfy a certain level of data protection requirements. However, these seals and certifications have a shelf-life of three years and can be revoked. This means that the organizations should proactively subject their compliance approaches to continuous monitoring and reviewing.

### 3.2.2. A Risk-Based Approach

The measures an organization needs to take in order to demonstrate compliance and good will should be proportionate to the risk each processing operation can impose on the people's rights. To evaluate the proportionality, then, the organizations are first encouraged to assess the potential risks that their processing activities might pose to users' fundamental rights. This is called a data processing impact assessment (DPIA). A DPIA will also show what measures one should take to prevent or minimize the risk of a privacy breach. DPIAs should be conducted before the beginning of data processing activities, however, they are mandatory only in high-risk situations (e.g., processing of special personal data, large scale data processing, the merging or combining of data gathered by various processes, use of newer technologies or biometric procedures, data transfer to countries outside the EU/EEC). Organizations should review the potential risks upon consultation with supervisory authorities called Data Protection Authorities, and continuously when adding new data processes or new technologies. A DPIA is therefore a dynamic requirement and can become mandatory upon adopting new technologies, processes and purposes. It is therefore a continuous process and requires organizations to keep monitoring the impact of their data processing activities in relation to the use of new technologies or establishing new processing purposes.

### 3.2.3. Data Protection by Design

ensuring compliance requires controllers to think about the impact of data processing operations on fundamental rights from the beginning, when designing these processes. It makes compliance easier because it enables the data controllers and processors to anticipate and discover privacy issues and tackle them at an early stage. To facilitate data protection by design, the GDPR has, for instance, repealed Directive 95/46/EC obligation on organizations to notify and register their data processing activities with local Data Processing Authorities. Instead, the GDPR requires organizations to maintain internal records of processing activities and be ready to make it available to a supervisory authority upon request. The idea is that, unless the organizations continuously keep a close eye on issues of data provenance and data cycles – including the reasons for which data is being collected and processed, where the data is stored at and how it moves in through and out of the organization – protecting personal data will turn into a more challenging issue. Thus, organizations are encouraged to keep records of the purpose of data processing, categories of data subjects and personal data, time limits for erasure of data, and where possible, a record of technical and organizational security measures. Unless such records are kept and reviewed, it would be more difficult to, for instance, demonstrate the lawful basis for processing (Article 6), or to set the conditions and requirements for obtaining consent (Article 7).

| General Data Protection Regulation (GDPR) ||
|---|---|
| **Problem** | **Instrument** |
| **Legal** ||
| ● **Lack of legal basis** for transferring data to outside the EU where data protection rules were not like those of Europe | ● Offering a modern toolbox of **various instruments for international data transfers**, including adequacy decisions adopted by the European Commission where the non-EU country offers an adequate level of protection, pre-approved (standard) contractual clauses, binding corporate rules, codes of conduct and certification. |
| ● **Outdated categorization of personal data** and the regulation not matching the new technological advancements | ● Updating the list of personal data categories and introducing new rules that take the risk of adopting new technologies into consideration. |
| **Technical** ||
| ● **Lack of technical preparedness** to understand the possible risks and appropriate measures to mitigate such risks | ● **Mandating** the assessment of the **impact** of **data processing activities** on individuals' rights for organizations that engage with high-risk categories of data and large-scale data processing. |
| • **Lack of privacy enhancement methods** to enable the **reuse** of personal data beyond the original purpose it was collected for | • Recommending **privacy-friendly techniques** such as pseudonymization and encryption |
| • The problem of **reuse of data for data categories** that were collected and processed without consent and based on other legal bases. | • **Separating data access** from **data portability** and **setting new rules** for data portability |
| **Economic** | |
| • **Red tape and associated costs** with notifying and recording data processing activities at data protection authorities. | • **Removing notifications and recording requirements** with data protection authorities. The organizations will maintain an internal record of data processing activities and will have to make it available to DPAs upon request. |
| • **Costs** associated with **data breach** notifications | • **Removing data breach notification**. Organizations will only have to notify Data Protection Authorities and individuals when the breach is likely to impact the data subjects |
| • **High costs of compliance** due to legal fragmentation across Member States | • **Legislating a unified Union-level instrument** that applies to all Member States |
| • **Difficulty and costly** provisions for enhancing **methods** of **recording data processing, conducting DPIA**s, and developing privacy-friendly techniques | • **Introducing** considerably **heavier fines for non-compliance** and negligence compared to the fines defined by Directive 95/46/EC<br>• **Taking active compliance measures into consideration when assigning fines** to organizations |

*Table 1. Summary of the EU Data Protection Regulation (GDPR)*

### 3.3. Interactions with Other Data-Oriented Legislation

The GDPR is important since all the discussed legislations in this article build upon it, and in case of contradiction, the GDPR prevails. Hence, anticipating the specific conditions under which the GDPR could interact with other legislation will have a substantial effect on the way organizations review and renew their governance approaches. For example, the GDPR interacts with the EU Data Act in relation to the processing and sharing of various data categories. The GDPR regulates one category of data (personal) and the Data Act regulates another (IoT data). Where the GDPR offers six different legal bases for collecting and processing data, the Data Act only allows for two legal bases. Thus, there are potential contradictions and tensions among different data-oriented legislations with the GDPR and with each other. This matters to vehicle manufacturers because they collect and process all sorts of different types of data - both personal and IoT - and it is critical that they understand these legislations interact and the different opportunities that different legislations afford. Next, as we present each piece of legislation independently, we will conclude with a reflection on their interactions and the opportunities that they make available.

# 4. EU Data Act

Like any major power, the EU wants global sovereignty and competitiveness in the digital data landscape. Its strategy is to create a single market for data and establish open and common data spaces where data from a number of strategic fields such as health, agriculture, mobility, manufacturing, energy, finance, and public administration can flow freely. The strategy ensures more data is available for innovative and sustainable product and service development in these sectors while keeping data generators in check. As a result, the EU has devised three major legislative instruments as part of its greater plan for Europe's single digital economy, the EU Digital Markets Act, the EU Data Governance Act, and the EU Data Act. These legislative instruments aim to regulate the market influence and flow of data by dominant actors (the so-called gatekeepers explained in Section 6), the flow of data held by public sector bodies (explained in section 7), and the flow of Internet of Things (IoT) data (explained below) held by manufacturers of connected devices and providers of related services. We first discuss the EU Data Act, the latest regulation of these three, since this regulation clearly and directly affects vehicle manufacturers, and it is the act that we have found in our research most concerns vehicle manufacturers.

The EU Data Act regulates the collection, processing, and sharing of IoT data generated through the use of connected devices and their related services. One example is the data generated through the use of vehicles connected to road infrastructure and users' phones. Such IoT data includes both personal (e.g., name, address, location, individual driving patterns, biometrics) and non-personal data including data related to the vehicles' operation captured by the sensors (EU Data Act, Recital 35). The data categories in scope of the EU Data Act are therefore more expansive and surpass personal data regulated under the GDPR.

## 4.1. The EU Data Act and Database Directive

The EU Data Act is the first comprehensive European legislative instrument that regulates the processing and sharing of industrial IoT data with the goal of strengthening users' rights on fair access and reuse of data. The Data Act regulates IoT data just as GDPR regulates personal data, and the two are often complementary. For instance, the Data Act complements the portability right under Article 20 GDPR which regulates personal data, by regulating the users' right to access and share IoT data that the users have a role in generating through the use of connected devices and related services. The Data Act amends other predecessor legislative instruments that aimed at protecting users' rights from different perspectives[6]. One of the most important issues that has been discussed, during the impact assessments[7] and calls for public consultations by the European Commission prior to the passing of the Data Act, is the copyright regarding (IoT) databases.

Databases are generally protected as "original" compilations by copyright law under the U.S. Copyright Act (Article 101), and European Database Directive[8] (Article 3). The U.S. Copyright Act defines an original compilation as a "collection and assembling of preexisting materials or of data that are selected in such a way that the resulting work as a whole constitutes an original work of authorship."[9] Similarly, the EU Database Directive defines original databases as "databases which, by reason of the selection or arrangement of their contents, constitute the author's own intellectual creation" (Article 3). Thus, existing protections focus on originality.

Besides originality in compiling data to merit protection, there is another argument that emphasizes the "substantial investment" in obtaining, verification, or presentation of the data in a compilation to also warrant protection. This argument has led EU's regulation on sui generis database protection laws for protecting databases the compilation of which have required substantial investments (Article 7, Database

---

[6] For instance, see Directive 2009/22/EC of the European Parliament and of the Council of 23 April 2009 on injunctions for the protection of consumers' interests.

[7] European commission Summary report on the Public Consultation on Data Act and Amended Rules on the Legal Protection Of Databases, 2021.

[8] Directive 96/9/EC of the European Parliament and of the Council of 11 March 1996 on the legal protection of databases.

[9] 17 U.S.C. 101.

Directive). Whereas the originality of databases has been subject to protection on a universal level, the sui generis rights have mostly existed in the EU. However, prior to the EU Data Act, there was a need to re-evaluate the applicability and usefulness of the sui generis database rights in the case of IoT data. In the evaluations of the Database Directive,[10] the EU commission identified several major objectives for regulating the sui generis database rights prior to the Data Act:

- **Harmonization effect on Member States legal approaches**: before the adoption of the Database Directive in 1996, Member States practices of protecting databases varied substantially. For instance, countries with common law tradition (e.g., UK) focused on the effort and resources that went into the creation of databases to set policies for database protection. Continental European countries emphasized originality and the intellectual creativity of the database. This is while Nordic countries focused more on the economic and technical efforts required to create databases. Therefore, non-original but resource-intensive databases enjoyed greater protection only in limited parts of the then European Community.

- **Economic impact to incentivize production of more databases:** the commission argued that protecting resource and investment intensive databases could provide incentive to improve the global competitiveness of the European database industry, in particular to close the gap in database production between the EU and the US.

- **Encourage data sharing incentives by safeguarding the balance between the interests of database producers and users:** the Database Directive would contribute to creation and use of databases (and thus contribute to database competitiveness and free information flow) by regulating when and what databases could be accessed. Examples include, a) protecting the rights of the *producers*: by allowing only "authorized" users to access "non-substantial parts of the database (Article 5), and b) protecting the rights of the *users* by recognizing exceptions such as use of databases for teaching and research, or for public security (Article 6).

- **Different economic and technical efforts required to create databases:** the examples of economic and technical efforts that the Nordic countries had in mind included the static and offline databases of the 1990s including company catalogs on CD-ROMs, or scientific and legal databases. The Nordic countries had thus generously extended their "catalog rules" to protect the more modern digital databases. This is while the way data can be generated and collected has significantly changed since the 1990s and many of these rules (e.g., the Nordic catalog rules) must be reviewed in light of the current economic and technical efforts required to generate and store data.

The Database Directive has been subject to three successive impact assessments prior to the passing of the EU Data Act. These assessments include, 1) the 2005 Database Directive Impact Assessment, 2) the 2018 Database Directive Impact Assessment, and 3) the 2021 Public Consultation on Data Act and Amended Rules on the Legal Protection of Databases.

Both the 2005 and 2018 Impact Assessments confirmed that the directive has effectively achieved its harmonization goals. However, they concluded that the sui generis right had 'no proven impact in the production of databases. Similarly, they showed that the sui generis right is instrumental to stakeholders mostly as a means of protecting databases against third parties, and not the stimulation of investment per se. Additionally, the assessments showed that the framers of the Directive took into account the economic and technical reality of the early 1990s, when the typical database of the time was static and offline (e.g., databases in publishers' industry). Whereas the assessment in 2005 and 2018 focused on the effects of the goals of Database Directive per se (i.e., impacts on legal harmonization, economic incentives, and free flow of data), the 2021 impact assessment evaluated the effectiveness of the Directive in relation to sharing IoT data which, at that time, was being regulated under the EU Data Act.

In the 2021-Public Consultation on "Data Act and Amended Rules on the Legal Protection of Databases," the majority (54%) of the respondents agreed that the 'sui generis' right should be reviewed, in particular in relation to the status of machine-generated data. Several reasons constitute the motivation to review the sui generis rights specifically in relation to machine generated data. For instance:

---

[10] Staff working document and executive summary on the evaluation of the Directive 96/9/EC on the legal protection of databases.

- The sui generis protection only prevents extraction and/or re-utilization of the entire database or a substantial part of the contents of that database. Extraction and reuse of the non-substantial part of a database are not subject to sui generis rights. In most of the cases, the requests are only for partial access to and reuse of the database.

- Additionally, the sui generis data protection right only applies to preserving the structure of the database not its content. In many industries (particularly in vehicle manufacturing), the structure of databases is common and developed based on the industry's best-practices. Therefore, there has been a lack of clarity regarding when exactly the 'sui generis' is effectively protective.

- Finally, the sui generis rights do not apply to databases which are the by-products of the main activities of an economic undertaking. In other words, spin-off databases are, in principle, not protected by the sui generis right, as they would not fulfill the 'substantial investment' threshold. Machine generated data in the case of IoT databases is arguably an example of this case.

These ongoing debates finally resulted in disapplying the sui generis rights to machine generated data in the Data Act. It should be noted that, the sui generis protection under Article 7 of the Database Directive does not specifically cover machine generated or IoT databases. The article provides a general protection for all sorts of databases that fulfill the "substantial investment" criterion. Under the Data Act, the legal grounds for such protection are explicitly nullified. Thus, the failure of the Database Directive over decades to demonstrate a positive economic impact or data sharing incentives has led the new regulations[11] to, in effect, *disapply* the sui generis rights.

### 4.2. Summary of the EU Data Act

The EU Data Act requires the owners and developers of connected products and providers of their related services to share data with their users and any third party of their choosing upon request. Of the three regulations mentioned above, the EU Data Act (hereafter Data Act) is the most demanding regulation for vehicle manufacturers with large amounts of data created through the use of connected devices and their related services (cars connected to mobile apps or electric chargers). Traditionally, this type of data has been harvested exclusively by manufacturers. The Data Act directly applies to the manufacturer of connected products and providers of related services. The Data Act requires vehicle manufacturers to share the data generated through the use of these connected products and services, with the users and any third party of their choosing (e.g., repair workshop, insurance, or banks). It also regulates the sharing of private sector's data with the public sector bodies to use in the performance of their statutory duties in the public interest. On a general level, the Data Act aims to:

– Enable public sector bodies to access and use data held by the private sector to help respond to public emergencies.

– Protect European businesses from unfair contractual terms.

– Facilitate access to and the use of data by consumers and businesses, while incentivizing businesses to continue investing in high-quality data generation.

– Allow customers to switch seamlessly (and eventually free of charge) between different cloud providers.

The data in scope of the Data Act ranges from any raw data generated automatically and without any further form of processing to data which have been pre-processed. Examples of raw data include data generated automatically by a single sensor or a group of sensors, or recorded by embedded applications, (e.g., data indicating hardware status and malfunctions). Pre-processed data is raw data that has been organized, structured, or analyzed for the purpose of making them understandable and usable prior to subsequent processing and analysis.

---

[11] A similar provision is also implemented in the Data Governance Act Article 5(7).

The range of data in scope is comprehensive and requires that data sharing take place quickly and seamlessly. This results in a process that the head of data strategy and business at a global vehicle manufacturer described as, "live data sharing." The problem is the many silo solutions, lack of technological infrastructure for live data sharing, and lack of economic assessments to be compliant with data-related requirements at this scale.

### 4.2.1. Silo Solutions and Individual Contracts

Contemporary data management and governance is new to many industries that have historically processed and analyzed digital data in limited ways. In the automotive industry, for instance, there has been a difference between the car manufacturer as the data controller and tier-one suppliers of digital services as data processors. Traditionally, data controllers have had a propensity to share as little data as possible with the processors under strict conditions and through individual contracts. This limited data sharing has occurred through silo solutions (to organize, process and share data) and based on several consents of a single product/service user. Various internal teams and delivery mechanisms at the OEMs need to be tied together to handle even such limited data sharing requests. An executive data strategist in one of the vehicle manufacturers we study painted a detailed picture of the data silo problem:

> *"If we've technically developed a solution whereby, we collect the data — like the autometer's status on how far the vehicle has traveled to your mobile app — per your consent, then we've traditionally made that in a siloed and locked solution. Now, there are a number of teams internally, and a number of different delivery mechanisms that need to be tied together <u>based on several consent of yours</u> for us to protect your data according to the GDPR. And now, we're supposed to use the same data and send it to the insurance company or a workshop of your choosing. Maybe that does not sound very complex, but in a big environment like ours, and with the thousands of applications and systems, and the complexity of the vehicle with hundreds of control units, it becomes very complex very soon.*

### 4.2.2. Technological Infrastructure

It is not just reorganizing the databases and rethinking consent approaches that will be onerous. To make the data readable and understandable for the users' designated third party, extensive investments should be made for improving interoperability and portability capabilities. That is, the ability to access and process data from multiple sources without losing meaning and then integrating that data for mapping, visualization, and other forms of representation and analysis. Yet, appropriate technical tools and standards are still being developed. As a vehicle manufacturer's data management enterprise architect exemplified, the systems that keep track of vehicle related data are not limited to the manufacturers' organizational boundaries:

> *Person A Sells the vehicle to person B. There's no obligation for Person A to inform the automotive company about the switched ownership? I mean, in Sweden, you need to inform the traffic administration that the registration name for the vehicle will be under the new owner's name. But there's no obligation to feedback that information to our organization. We need to have a way to record that data. Actually, we already have a DMA [data management agreement]. It's a database to keep track of this information but the coverage is so bad, and we are discussing how to improve this type of data management.*

The information about the change of vehicle ownership is usually recorded within some national authorities and is not necessarily communicated back to the manufacturers. This creates problems for correctly identifying the vehicle owner who has the right to IoT data access. The example highlights that the data-related infrastructure needs to be managed by not only the vehicle manufacturers but also by Member States. However, since vehicle manufacturers have historically shared IoT data on a limited scale, the current data management infrastructure does not match the requirements for complying with the scope and speed of data sharing at the focus of Data Act.

### 4.2.3. Lack of Economic Assessment

Vehicle manufacturers and their suppliers have spent years and billions of Euros to develop digital devices that generate data for monetization. The business models around data analytics are quite nascent. For many vehicle manufacturers, there is no data analytics culture in place, the techniques to capture high quality data are still developing, and the data-driven projects and initiative are still scant and limited to a few stand-alone projects that do not mount to a solid digital business model. Despite being new in exploring the potential of data, OEMs are now required to share it with competitors who have not made corresponding investments. This will arguably give competitors unfair advantage.

The compensation for making data available is regulated under Article 9 of the Data Act. It, for instance, emphasizes that several issues – including but not limited to costs incurred in making the data available, (e.g., formatting, dissemination via electronic means and storage), parties who contributed to obtaining, generating or collecting the data, and the volume, format and nature of data – should be taken into consideration when calculating the costs and assigning the compensation fees. However, manufacturers have not typically been required to conduct such economic assessments and most of the databases (e.g., extended vehicle platforms) were subject to IP, and sui generis, or trade secret rights. Some of these rights are disapplied by the Data Act.

One example is the sui generis rights in the Database Directive[12]. To comply with the Data Act, the sui generis rights that applied to preserving databases for which "substantial investment" had been put in collecting, organizing and storing the data (Article 7), is disapplied to the data captured through physical components (e.g., sensors) and through the use of connected devices and related services (Data Act, Article 43 and Recital 112). Since such data is generated automatically by the sensors and through the use of the devices and services, one interpretation is that no "substantial investment or effort" has been put into developing databases that capture such data. Therefore, some economic valuations of pre-DA times may not be valid any longer. However, creating new data management and storage structures (that can, for instance, capture the necessary metadata and connect it to the right data categories) will bear financial costs. It is unclear whether these costs could be counted for in compensation fees for procuring data to third parties under article 9 of the Data Act. The disapplication of the sui generis rights does not so much impact maintaining the existing databases as it will impact economically the creation of *new* data catalogs.

The most important long-standing problems in data sharing and the corresponding Data Act provisions that address those problems are presented in the table below. They include overarching gaps in the legal process (contractual terms and obligations, data in and out of scope), technical capabilities (data silos, poor data management, interoperability capabilities), and economic incentives (regulating trade secrets, IP and database, as well as competition rules).

| EU Data Act (DA) ||
|---|---|
| Problem | Instrument |
| **Legal** ||
| • **Absence** of clear **rules** on **data types in scope** of mandatory sharing. | • **Clarify** the types of data in scope |
| • **Abuse** of **contractual imbalance** (imposing unilateral contractual conditions). | • **Impose conditions to safeguard** enterprises from **unjust**, or **unilateral contractual terms** imposed by a party with a considerably stronger market position. |
| • **Lack of legal basis for businesses to share data with public sector bodies** | • **Imposing strict conditions** for when/where/how businesses must share or can reject the data access request by the public sector bodies (does not hold true for microenterprises and small enterprises to avoid imposing unbearable burdens on them) |
| • **Unilateral decision powers** by the manufacturer about when/where/how data should be portable | • **Impose** a list of **minimum provisions** that must be included in **contracts** for the provision of data processing services |
| **Technical** ||

---

[12] Directive 96/9/EC of the European Parliament and of the Council of 11 March 1996 on the legal protection of databases.

| | |
|---|---|
| ● **Absence** of **standards** ensuring **semantic and technical interoperability** | ● **Encourage** participants that offer data or data-based services to other participants within and across common European data spaces to **support interoperability of tools for data sharing** including **smart contracts.** |
| ● **Absence** of **universal authentication** methods | ● **Emphasizing** the **prevalence** of the GDPR and national laws when the data requested conflicts with privacy rights on personal data. |
| ● **High** level of **fragmentation** of information in **data silos** | ● **Regulate** private sector's data sharing bases on a **Union Level** |

| Economic ||
|---|---|
| ● **Poor metadata management** | ● **Mandating** the **sharing of metadata** necessary to interpret and use IoT generated data, to a third party without undue delay, of the same quality as is available to the data holder, easily, securely, **free of charge to the user**, in a **comprehensive**, **structured**, **commonly used** and **machine-readable format** and, where relevant and technically feasible, **continuously and in real-time** |
| ● Lack of regulations to protect **Intellectual Property**, **unique databases**, and **trade secrets** | ● **Reviewing** certain aspects of the Database Directive, particularly focusing on elucidating the role of the **sui generis database rights.** |
| ● **Lack of mechanisms to protect the benefits and incentives of data generators to continue investing in high-quality data generation** | ● **Excluding** the designated **gatekeeper** as eligible third parties.<br>● **Prohibiting** data requesters to use the data for creating competing products and services.<br>● **Prohibiting** the designated gatekeepers to solicit or commercially incentivize a user, in any manner, to make data available to one of its services that the user has obtained |
| ● **Unreasonable prices for data procurement** | ● **Establishing a monitoring mechanism on switching charges** imposed by providers of data processing services on the market, and to further specify the essential requirements in respect of interoperability |
| ● **Costs** of **contracting** and implementing **technical interfaces** | ● **Conferring implementing powers to the Commission** in respect of the **adoption of common specifications** to **ensure the interoperability** of data, of data sharing mechanisms and services, as well as of common European data spaces and the interoperability of **smart contracts** |

*Table 2. Summary of the EU Data Act (DA)*

The Data Act requirements thus disrupt the data sharing climate to which vehicle manufacturers have grown accustomed, requiring them to rethink their legal and technical infrastructure. To help manufacturers comply with a seamless data sharing process, the EU commission has set up several frameworks to help with, for example, reviewing the rules on intellectual property, trade secret and sui generis database rights, and supporting the development of interoperability and portability capabilities, as well as more innovative contractual processes (e.g., smart contracts).

The Data Act, on its own, puts significant demands on vehicle manufacturers. But it is important to note that vehicle manufacturers need not only be compliant with the Data Act in isolation, but also to ensure that the new data sharing requirements do not clash with other directives and legislative instruments that they have operationalized based on cautious and limited data sharing. The following are a few examples.

### 4.3. The Data Act and GDPR

- **Processing and sharing different categories of data - consent and contract vs. other legal bases:** the Data Act regulates the collection, processing, and sharing of another data category, i.e., IoT data generated through the use of connected devices and their related services. One example is the data generated through the use of vehicles connected to road infrastructure and users' phones. Such IoT data includes both personal (e.g., name, address, location, individual driving patterns, biometrics) and non-personal data including data related to the vehicles' operation captured by the sensors (Data Act, Paragraph 35). The data categories in scope of the Data Act are therefore more expansive and surpass personal data regulated under the GDPR.

The challenge is that IoT data can include personal data, too. For instance, while the data holder's de facto position may be the sufficient legal ground (e.g., based on legitimate interest) for collecting and processing personal data, Article 4(13-14) of the Data Act clearly states that it is not sufficient for either processing or sharing of any non-personal data. Thus, both processing (e.g., to improve IoT devices) and sharing of IoT data is only allowed based on contractual agreements with users and with their direct consent. This overlap implies a significant legal change from the current situation (Kerber 2023), because it assumes that data holders have contractual agreements with users for processing all types of data. So far, these agreements or consents have only been necessary for sharing data not for processing them.

Additionally, it is worth noting that a customer's request for sharing their data with a third party does not equate to their consent. But the regulations are too general and do not provide clear guidelines for the terms and conditions of data sharing in the EU Data Act (Casolari et al. 2023). Since the GDPR always prevails in the case of complications, organizations might benefit from reviewing the consent and contractual terms for all these categories of data. They might be required to adopt approaches that not only help track personal and non-personal data categories and when/where they impose risks to privacy and other fundamental rights of users, (See Finck and Pallas 2021) but also allow the organization to easily track what data types are already tied to user consent or fulfillment of specific contractual terms.

- **Interfaces for reviewing contractual terms and consent - personal data vs. IoT data:** The right to withdraw consent under Article 7 of the GDPR gives data subjects the right to withdraw consent at any given time. The Data Act gives users control over other categories of data through consents and contracts for processing and sharing that can be withdrawn at any given time, too. This will affect the way businesses ensure compliance with both the GDPR and the Data Act in relation to reviewing, maintaining or revoking consent.

    One tension is related to conducting DPIAs. Vehicle manufacturers are subject to mandatory DPIA under the GDPR. Assessing risks of processing data depends on tracking the time span during which different categories of data (personal vs. IoT) are allowed/required to be maintained for different processing and sharing purposes before they pose any risks to users' privacy and fundamental rights. Different categories of data (e.g., IoT data generated through the use of connected devices and services vs. more static categories of data related to ledgers, accounts, assets, or address) imply different levels of continuity and connectivity and hence, control over and risks for users' rights (Politou et al. 2018).

    For users too, having control over their IoT data is different from personal data. It is difficult for users to understand the size and scope of their collected data (Rantos et al. 2019), and form a mental model of how complex functions, devices, organizations, and third parties will be connected to each other, for what exact/later purposes the data will be used (Williams et al. 2018). At the same time, the continuous, connected, and ubiquitous, and constantly evolving nature of IoT data renders the mechanisms of one-time single consents/contracts problematic. Emphasizing the importance of "informed" and "transparent" user control over their information, Articles 13 and 14 of the GDPR require data controllers to provide users with exact information about the specific use and purposes of collecting, processing and sharing data and provides them with the right to lodge a complaint against the collecting, processing, and sharing of their data.

    Thus, vehicle manufacturers not only need to track the risks of processing/sharing such data continuously – to comply with DPIA requirements – they also need to review the design of the interfaces (Williams et al. 2018) that allow users to stay informed about IoT data processing and sharing processes (Neisse et al. 2016) continually – to avoid possible risks and complaints.

- **User identification requirements - no need to identify users in processing personal data vs. the need to identify users for sharing IoT data:** There is a tension between the Data Act and the GDPR in relation to the need to identify individuals. The idiosyncrasies of IoT data from connected devices go beyond its sheer volume. IoT data has high specificity and continuity (Oberländer et al. 2018). IoT data has the capability to yield a continuous flow of diverse and revealing information taken from multiple information points (Cichy et al. 2021). Some examples include both personal and non-personal data such as data on location, biometrics, health data, acceleration and braking, frequency of refueling and of tire pressure adjustments. This diversity of data points together with the continuity of data flow affect not only privacy issues for users but compliance issues for manufacturers. To comply with the Data Act, manufacturers need to employ proper means to identify the users correctly to be able to share the IoT data with the right individual/user. This creates challenges for vehicle manufacturers when they process IoT data which might not be tied to the user's identity. Under Article 11 GDPR, if the purposes for which a controller processes personal data do not or do no longer require the identification of a data subject, the controller shall not be obliged to maintain, acquire or process additional information in order to identify the data subject to stay compliant with the GDPR (e.g., data portability rights).

To illustrate the concern, consider users of heavy-duty vehicles in the context of industrial IoTs (IIoTs), where multiple competing data generators and consumers exist in different stages of the system (Zheng et al. 2020). The data is generated by multiple businesses in the logistics and freight chains including manufacturers (e.g., chassis makers, and bodybuilders), multiple drivers who drive the same vehicle at different points of time, or various companies that lease a fleet of heavy vehicles. The stream of data in this chain is naturally valuable to several parties and users that are constantly changing. Along this chain of alternating users and stakeholders, the vehicle manufacturer might not need to process personal data for purposes that no longer require the manufacturer to identify a data subject (e.g., change of the user, or the driver). Then the manufacturer is not required to either seek/maintain further information to identify the data subject, or to give the data subject access and right to portability of their data (Article 11 (1-2) GDPR). For instance, consider the case when the driver has changed, and the purpose of processing the personal data of the previous driver (e.g., analyzing human driver behavior under certain operational conditions of the vehicle to improve system safety) is not tied to the identity of the driver. This case can be problematic when the vehicle manufacturer is to comply with the Data Act and provide the user (be it the driver or the freight company who leases the vehicle and employs the driver) with the data generated through their use of the connected vehicle without having the means to identify the driver. As the technical manager of a vehicle manufacturer mentioned:

*"It's quite an overwhelming situation with all the overlaps between various legislations. A major issue is to confirm who the user is tied to the data. The users are constantly changing, and we need to keep updating that info over time and over multiple nations. I mean even if it is still in the EU, we don't have a common way of confirming who's who. The user can also be a private person or a legal entity [which imply different privacy assessment requirements].*

### 4.4. Interaction with Future Amendments to the EU Regulation on Type Approval: Opportunity

- **Monetize industry-specific standards to share in-vehicle data directly with third parties: Proactive approach vs. reactive compliance:** Second, another instance regards the multiple existing standards that regulate external access to in-vehicle data (e.g., UN regulation on automated lane keeping systems). These safety and security critical standards require manufacturers to comply with a strict set of criteria such as audits, type approvals, and documentations. One of the areas that is directly affected by providing access to in-vehicle data such as the data that comes directly from the sensors and vehicle safety components is conducting comprehensive risk assessments mandated both by the GDPR and AI Act (explained in section 7) to ensure safety, security and privacy compliance. At the same time, all car data must be cleansed, normalized, and harmonized before it is usable. The *Type Approval Regulation*[13] calls for open data formats and well-defined metadata standards not fully developed yet. All these assessments, audits, and technical processes create extra tasks and costs for vehicle manufacturers.

Due to increased security and safety breaches, OEMs have provided access to this type of data via the extended vehicle platforms which, in turn, have bestowed them with a dominant position to control what data is being accessed by third parties. The majority of after-market product and service developers consider manufacturers' proprietary control over the data in the extended vehicle platform to negatively affect fair market competition (Gill 2022; McCarthy et al. 2017). For instance, the aftermarket product and service providers argue that much of the essential repair and maintenance services depend on remote access to in-vehicle data and over-the-air-updates that are not even stored in the extended vehicle platforms and are therefore not shareable according to the Data Act. Instead, these after-market developers propose that On-Board Application Platforms should replace extended vehicle platforms. On-Board Application Platforms can host 3rd party applications that directly access vehicle data. This ongoing debate is pushing the Type Approval regulators to consider these technological advancements. In fact, as a lead data policy officer at the European Commission mentioned, some policy makers are also arguing for extending the data categories in scope of the Data Act to include in-vehicle data sharing as well. Thus, several regulatory proposals are already circulating in the European Commission to accommodate the after-market developers' ambitions of accessing in-vehicle data.

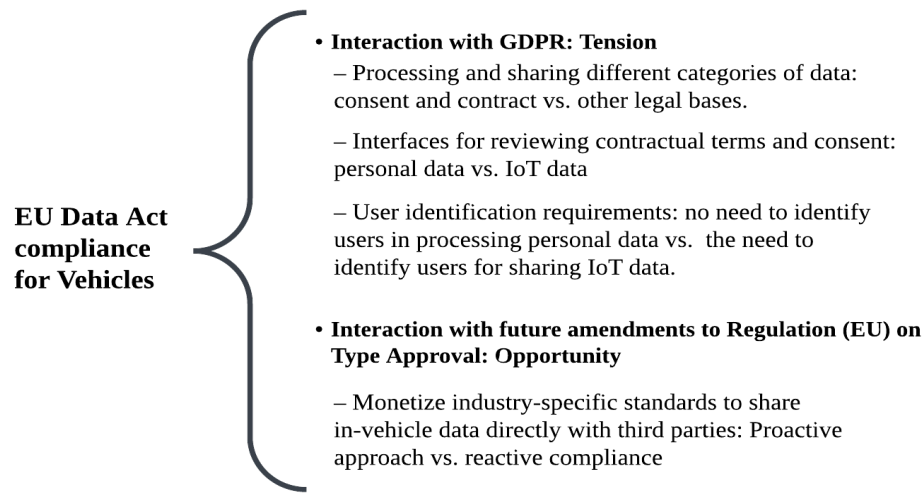

*Figure 2. The Interaction of EU Data Act with Other Regulations*

According to the Data Act (Recital 15), the data sharing requirements are applicable when a) the data captured by the applications and systems during the use time is retrievable by the manufacturer, and b) when capturing such data does not require manufacturers to use complex algorithms and processing techniques that could be subject to IP laws. Thus, granting access to in-vehicle data is not mandatory at the moment since the raw data is not stored and retrievable by the manufacturer. However, as the lead data policy officer at the European Commission emphasized, right at this point, one value creation potential for vehicle manufacturers emerges. Vehicle manufacturers can adopt a proactive approach toward granting access to in-vehicle data through developing APIs and proper standardizations, as well as interoperability capabilities that allow third parties direct access to vehicle data in exchange for corresponding compensation according to Article 9 of the Data Act. In the proactive approach, for instance, the manufacturers have the opportunity to argue for considering the extra costs — for being compliant to the GDPR requirements regarding assessing the impact of processing and sharing new categories of data (i.e. in-vehicle data), as well as the additional risk assessments required to ensure safety of AI systems regulated under the EU AI Act (see section 7) — in their compensation schemes. This is while, if vehicle manufacturers await regulatory mandates for sharing in-vehicle data in future, they might not be able to charge compensation fees. This is because any data category that eventually becomes shareable with users through regulatory mandates can be shared by users with the after-market (third) parties beyond the manufacturers' control and with little financial compensation.

---

[13] Regulation (EU) 2018/858 of the European Parliament and of the Council of 30 May 2018 on the approval and market surveillance of motor vehicles and their trailers, and of systems, components and separate technical units intended for such vehicles, amending Regulations (EC) No 715/2007 and (EC) No 595/2009 and repealing Directive 2007/46/EC.

# 5. Data Governance Act

Whereas the Data Act is about *mandatory* data sharing by the *private* sector (e.g., businesses), the EU Data Governance Act is all about encouraging *voluntary* data sharing by the *public* sector bodies (e.g., traffic administration, healthcare system, governmental authorities).

## 5.1. The Data Governance Act and Open Data Directive

Historically, the European Union has worked toward facilitating the sharing of public sector's data with two important objectives. First, the EU has argued that the freedom of information and open data initiatives will lead to more transparency of the inner workings of the public sector bodies (open government concept) and therefore increased participatory and collaborative dialogue between citizen participation (social impact). The second incentive is based on the argument that the reuse of important datasets held by the public sector bodies will have invaluable economic gains (the economic impact). Both incentives are interlinked and have received primary and secondary attention.

For instance, with regards to the reuse of public sector information, people first took notice of the large amounts of data held by public sector bodies — and the potential lying in the reuse of such data — in the 1970s. The reuse of public sector information was first officially recommended around the 1990s in the form of non-binding guidelines which set the bare minimum uniform rules on key issues such as pricing and licensing and encouraged open data policies and open data formats. Then the first European directive on public sector information (PSI), i.e., the PSI Directive, formed in 2003[14] to harmonize and more firmly facilitate the reuse of public sector information. The PSI Directive was then amended in 2013 (10 years later) to match up with the technological evolution as well as EU Commission's Digital Agenda.

Meanwhile, the open data movement dates back to the year 2000, when non-state actors pressured the government to proactively share non-sensitive data — as a cornerstone of democracy and government transparency — and make it available for reuse without any legal or technical restrictions. This movement continued to make its impact on the PSI Directive of 2013[15] (and 2014) in the form of technical requirements for open data formats and recommended standard licenses[16].

By the end of the decade, however, the role of data had become so crucial to the European economy that, rather than emphasizing the reuse of data, the EU was advocating open data spaces and single digital markets as the agenda for the future. Thus, the Open Data Directive[17] adopted and replaced the PSI Directive in 2019 and embraced the EU's open data agenda in its title. It must be noted that granting access to data is within the jurisdiction and authority of Member States and the Open Data Directive recognizes this restriction (Article 1(2)). However, it has notable obligations for making data available as long as it does not require disproportionate efforts.

The fast pace of technological evolution and data generation required assessing the impact of the Open Data Directive quickly. The 2020 Impact Assessment[18] revealed several major problems on the way for Europe to actualize its Open Data and Single Digital Market goals:

1. **Fragmented approaches and lack of a data exchange culture in the public sector:** as will be explained in the next section, the EU law bestows the authority upon Member States to regulate access to data. Different Member States have, over decades, shown different levels of openness to granting access to public sector data. At the same time, public sector bodies have had difficulties in managing risks related to the reuse of data, especially personal data. Although the GDPR has raised awareness[19] about protecting personal data, the awareness has not necessarily led to increased competence and expertise to protect data[20].

---

[14] Directive 2003/98/EC of the European Parliament and of the Council of 17 November 2003 on the re-use of public sector information.

[15] Directive 2013/37/EU of the European Parliament and of the Council of 26 June 2013 amending Directive 2003/98/EC on the re-use of public sector information.

[16] This short historical account provided here is adapted from Broomfield 2023. For a more detailed and synergistic account of the origins and evolution of "open data" and "secondary use of public sector information", as two interlinked but separate concepts, see (Broomfield 2023).

[17] Directive (EU) 2019/1024 of the European Parliament and of the Council of 20 June 2019 on open data and the re-use of public sector information (recast).

[18] Commission Staff Working Document Impact Assessment Report Accompanying the Document Proposal For A Regulation of the European Parliament and of The Council on European Data Governance (Data Governance Act), 2020, Brussels, 25.11.2020, SWD(2020) 295 final.

2. **Limited data categories in scope of the Open Data Directive:** the Open Data Directive regulates the reuse of non-personal and nonsensitive data. This is while high value datasets held by the public sector are usually protected by intellectual property, trade secrets or confidentiality rights. Examples include data held by the finance, agriculture, health, energy and transport government bodies.

3. **Platformization of data exchange:** In the absence of harmonized regulations and fragmented approaches across Europe, one form of data brokers was taking over the market, i.e., the Big Tech companies who with their technical and network capabilities dominated the entire data-driven value chain in collecting, storing and processing large volumes of data at their disposal and growing. As intermediaries of data collection, storage and processing, the Big Tech companies have been absorbing more market power and dominance over other players to whom they provide any of the mentioned services.

4. **Low trust in data sharing business models:** Without an alternative European model for data exchange intermediation, many industry players express distrust in existing integrated tech service providers as platforms for industrial data exchange. Offering an alternative intermediation model (that is neutral and heavily regulated) is, therefore, one of the key amendments in the EU Data Governance Act (2023) which complements the Open Data Directive and addresses the above-mentioned shortcomings.

The EU Data Governance Act requires the creation of European Common Data Spaces to make highly valuable datasets findable and accessible. This key issue is applauded by many stakeholders (in the impact assessment of the Open data Directive) who believe that data spaces operate as an exemplary case to clarify control rights and rules on data access and use, particularly in relation to data of sensitive and protected nature.

### 5.2. Summary of the Data Governance Act

The EU Data Governance Act is all about encouraging voluntary data sharing by the public bodies across different sectors (e.g., traffic administration, healthcare system, governmental authorities). Public sector bodies rely on vast amounts of data. However, they do so not for business gains but to improve their operations and services in the interest of the public. The problem is that a large amount of data held by the public sector includes protected or sensitive data such as individuals' health or criminal records, geographical coordinates, land registry, statistical or legal information. At the same time, the European Member States have cultivated different sentiments toward transparency and therefore comfort in sharing such data, with Nordic Member States, for instance, being more enthusiastic, and Member States such as Romania or Netherlands being more reserved. Yet, even in the Nordic countries, the public sector's data is handled carefully and conservatively. This has resulted in different strategies to store and grant access to the public sector's data across different Member States including different formats and modes or degrees of openness.

The Data Governance Act is part of the EU's package of measures to promote a single EU market and a data economy that includes a free flow of data and the development of artificial intelligence. It complements the EU's Open Data Directive[21] that was devised with the same goals as the Data Governance Act. However, since the Open Data Directive was being enforced on the Member State level, the free flow of information across sectors and member States was impeded by individual Member State laws and strategies.

---

[19] Communication from the Commission to the European Parliament and the Council Data protection as a pillar of citizens' empowerment and the EU's approach to the digital transition - two years of application of the General Data Protection Regulation.

[20] Commission Staff Working Document Impact Assessment Report Accompanying the document Proposal for a Regulation of the European Parliament and of the Council on European Data Governance (Data Governance Act).

[21] Open Data Directive

Yet, the Open Data Directive set the most important basis for the Data Governance Act to be effective. The basis includes mandating the presentation of data in both open and dynamic data formats. Open format refers to formats that are machine readable, accessible, and findable through electronic means and include metadata. Dynamic data refers to documents in a digital form, that are subject to frequent or real-time updates. The Open Data Directive however did not cover measures for several important issues covered in the Data Governance Act: a) legal and technical infrastructure to protect sensitive data including commercially sensitive data, data subject to confidentiality requirements, data protected by intellectual property rights, and individuals' personal data, and b) substantially reducing data procurement fees, and c) promoting the voluntary sharing of personal data for altruistic purposes. These areas are regulated under the Data Governance Act at a Union Level which harmonizes the approach as well as legal and technical infrastructures to promote a free flow of data across sectors and Member States. The Data Governance Act aims, on a general level, to:

- Increase access to public-sector data for the purpose of developing new products and services (commercial) and research initiatives and altruistic purposes (non-commercial).

- Develop the legal base at the Union level to promote public's trust in sharing sensitive data.

- Develop the technical infrastructure and methods that help share protected data in a secure and trustworthy way.

It is the European Union's position that data which has been generated or collected by public sector bodies, at the expense of people's budget, should be available to and benefit the people (Recital 6). It is also the Union's stance that the main culprits for withholding data sharing and processing in this sector have been a) the high sensitivity of the data types, and b) a low trust in the technical and legal procedures for reuse of such data. The ambition is to do so by means of three major instruments:

- **Data intermediation services**: the EU aims at fostering a new business model – data intermediation services (hereafter DIS). DIS providers will function as neutral third parties and connect individuals and companies with data users and provide a trusted and secure environment in which they can share data. Intermediaries can help organizations to lower the costs and knowledge required to share data, particularly if building this infrastructure is prohibitively costly for data providers or users. For instance, DIS providers can supply the organizations with embedding effective cybersecurity measures, or apply privacy-enhancing technologies (PETs). They can also build bridges between different data providers enhancing interoperability and data portability. Finally, they can create trusted research environments for research and innovation. Fostering the DIS providers addresses the lacking technical procedures previously lacking.

- **Regulating DIS providers and enforcing the regulations on Member States**: The Data Governance Act is largely about voluntary data sharing. The mandatory section mostly regards the DIS providers. In other words, regulating DIS providers across all Member States is the reason a Union level regulation such as the Data Governance Act was required. According to the Act, the DIS providers are to comply with strict requirements to guarantee their independence and neutrality towards the parties that are exchanging data. They must ensure a strict separation between the DIS and any other services they provide to customers. It also prohibits the DIS providers to use the data exchanged via their data-sharing platform for their own purposes (Chapter III, Article 10-12). The Member States are mandated to fully enforce these regulations and punish violations. Regulating the intermediaries addresses the legal procedures previously lacking.

- **Single digital gateway for data sharing**: to ensure that the intermediaries can access and complete the procedures of requesting data and processing these requests fully online and in a cross-border, transparent and non-discriminatory manner, the Data Governance Act mandates the creation of a single digital gateway both at the Member State and the Union Level. The single digital gateway at the Member State level is called *The Single Information Point* (Chapter 2, Article 8) which will be competent to receive enquiries or requests for the re-use of data and to transmit them, where possible and appropriate by automated means. It will make available by electronic means a searchable asset list containing an overview of all available data resources including, across sectors, and regions, with relevant information describing the available data, including at least the data format and size and the conditions for their re-use. The single digital gateway at the Union level is called the *Common European Data Space* (Chapter I, Paragraph 2). This Union level entity logs all the information points available at the Single Information Points across Member States so that all the data communication and access will be digitized and harmonized.

| Data Governance Act (DGA) | |
|---|---|
| **Problem** | **Instrument** |
| **Legal** | |
| ● The gap in the Open Data Directive that **did not allow** the **re-use** of protected data such as **commercially sensitive data**, data subject to **confidentiality** requirements, data **protected by intellectual property right**s, and **individuals' personal data**. | ● Designate, confirm, regulate, and closely monitor **data altruism organizations** at the **Union level** as eligible to **handle personal and commercially sensitive data** strictly for **altruistic** purposes. |
| ● Exclusive access granted to specific parties based on individual contracts | ● Strictly regulate the data **access grants** and **denials** by mandating the public body sectors to **log** all **requests**, and **reasons** for granting or denying access to data.<br>● Prohibit exclusive data access grants. |
| **Technical** | |
| ● The general public's **low trus**t in the mechanisms for handling sensitive non-personal data | ● Designate, confirm, regulate, and closely monitor data intermediation service providers at the Union level as eligible to Develop the **technical infrastructure** and methods that help **share protected** data in a **secure and trustworthy** way. |
| ● Fragmented ways of **searching** and **accessing** data | ● Create **Single Information Points** at the **Member State** level and Common **European data spaces** at the **Union** level that make data findable, accessible, interoperable and re-usable (the '**FAIR data principles**'), while ensuring a high level of cybersecurity |
| **Economic** | |
| ● Questions about the **fees** regarding the **non-commercial** re-use of data held by the public sector bodies | ● **Encourage** the public sector bodies to allow reuse at a **discounted fee** or **free of charge**, for example for certain categories of re-use such as non-commercial re-use for **scientific research purposes,** or re-use by **SMEs** and **start-ups**, civil society, and educational establishment |
| ● Questions about the **fees** regarding the **commercial** re-use of data held by the public sector bodies | ● **Mandate** the **basis** by which **fees** are to be assigned based on complete transparency and non-discriminatory criteria visible and accessible for the public observation.<br>● **Limiting** the fees to **only necessary costs** for organizing, clearing and procuring data.<br>● Assign the **Member States** as the authority to **define** the basis for **fees** |

*Table 6. Summary of the EU Data Governance Act*

Vehicle manufacturers may think that the Data Governance Act is immaterial to them, and they are largely correct. However, there may be some benefits of the Data Governance Act for vehicle manufacturers. For example, vehicle manufacturers may benefit from encouraging government bodies to share data that is applicable to their organizations, perhaps through membership in consortia, and thus service their customers better. Further, the Data Governance Act may complement their approach to complying with other regulations such as the Data Act (see next section).

### 5.3. The Data Governance Act and Data Act: Opportunity

- **Cooperation contracts: data platformization vs. neutral intermediary business models:** data and tech heavy industries such as vehicle manufacturing and aviation rely on research consortia and collaborative platforms that can pool engineering and operations data in a secure ecosystem. Examples include Skywise and Otonomo. Skywise is a consortium in the aviation industry that connects in-flight, engineering, and operations data and is used by suppliers, as well as over 100 airlines. Otonomo is a vehicle manufacturing data exchange platform that combines data from connected cars and fleets located globally, and enables car manufacturers, drivers, and service providers to be a part of a connected ecosystem to create innovative apps and services. Over the years, these data intermediation platforms have commoditized data and capitalized on the possibilities for data collection and analysis engendered by networked communications and computing (Crain 2016). This platformized business model has disadvantages since it allows bigger players to pull more data toward themselves, offer more products and services and further centralize their power. Some scholars have referred to this centralization of power as information capitalization (Cohen 2019) and platform capitalism (Srnicek 2016). For instance, in recent years, different actors in the aviation industry have suggested that Skywise is, ultimately, part of a duopoly (Airbus and Boeing) in a competitive industry. Thus, the benefits of sharing data with Airbus must be assessed against granting too much power to this corporation. Such criticism of intermediation business models has turned into one of the main incentives for regulating data intermediation service providers.

The Data Governance Act seeks to create a more trustworthy and fair research and collaboration environment by preventing platform capitalism. It does so through two main mechanisms: 1) structural separation: unbundling of services which requires DIS providers to be structurally separated from any ancillary commercial activity, and 2) limitations on data use which prohibits DIS providers to use the data subject to intermediation services 'for purposes other than to put them at the disposal of data users" (Data Governance Act Article 12(a)). These regulations limit vertical and horizontal integration (Ditfurth and Lienemann 2022) and create contractual freedom for vehicle manufacturers to form research and development cooperations without fearing conflict of interest (Richter, 2023).

These legal instruments of the Data Governance Act align with some of the provisions in the Data Act that prohibit unfair data sharing terms and applications. For instance, the Data Act considers a contractual term to be unfair and therefore prohibited if its object or effect is to allow a party with more power to access and use the data of another party in a manner that is significantly detrimental to the legitimate interests of the other party, in particular when the data is commercially sensitive or is subject to trade secrets and intellectual property rights (Article 13, Paragraph 5(b)). An example of such a use could arguably be that the data recipient or a third party are prohibited from using the received data to develop a competing connected product (Data Act, Article 11, Paragraph 3 (b)). Additionally, the Data Act considers a data sharing contract to be unilaterally imposed if it has been supplied by one party and the other party has not been able to influence its content despite an attempt to negotiate it. The Data Act places the burden on the party that has supplied the contract to prove that the contract terms are not unilateral (Article 13, Paragraph 6).

Thus, the business model of neutral data intermediation service providers can eventually offer an alternative to proprietary platforms that enable sector-specific cooperations for research and development. Together with the protective measures of the Data Act mentioned above, the neutral DIS business model of the Data Governance Act could create a space for vehicle manufacturers to form new cooperation contracts and alliances.

### 5.4. The Data Governance Act and GDPR: Opportunity

- **Forming major inter-sectoral research consortia: setting the standards vs. adopting the standards:** Since the GDPR days, research and public interest causes have always merited specific recourse to access and process data which is otherwise protected. For instance, Article 9(2-j) and Article 89(2)) allow for processing special categories of data for public interest and Recital 159 of the GDPR also emphasizes that privately funded research for public interest is also subject to such exemptions. However, it has traditionally been difficult to motivate the access and processing of sensitive or protected data for private sector[22] actors due to lack of trustworthy protective mechanisms.

By designating trusted DSI providers, the Data Governance Act seeks to develop the trustworthy mechanisms and spaces where such data could be shared and processed among actors from both public and private sectors. These intermediaries invest in the time, processes, technology, and joint environments necessary to facilitate safe and secure data transactions. In exchange for their advanced technical capabilities, vehicle manufacturers can access important datasets held by relevant public bodies (e.g., traffic administrations, weather agencies and healthcare) for sectoral innovation. Access to data such as the city's traffic map can contribute to forming major research projects instead of small-scale fragmented research initiatives. The technical manager at one of the vehicle manufacturers in our study emphasized the importance of these potentially large-scale research projects in which different parties offer different capabilities. For instance, some stakeholders offer robust metadata management capabilities which have always been an impeding issue for successful data exchange[23], vehicle manufacturers can offer aggregated data which requires processing that might be subject to protection and therefore not in scope of the Data Act, and the traffic authorities can provide valuable datasets:

*We've been part of the research project EuroLog with the public sector and you know, within its test directory, we have this 'natural access points on the map' core project, for instance, where they're trying to create hubs where you can actually change the metadata about where to find data sources. That I think could be very valuable for many parties, if it becomes a little bit more successful. There, we can probably provide the aggregated data for such initiatives. And there's for sure data from traffic authorities and other stakeholders that could be used to provide really good valuable services.*

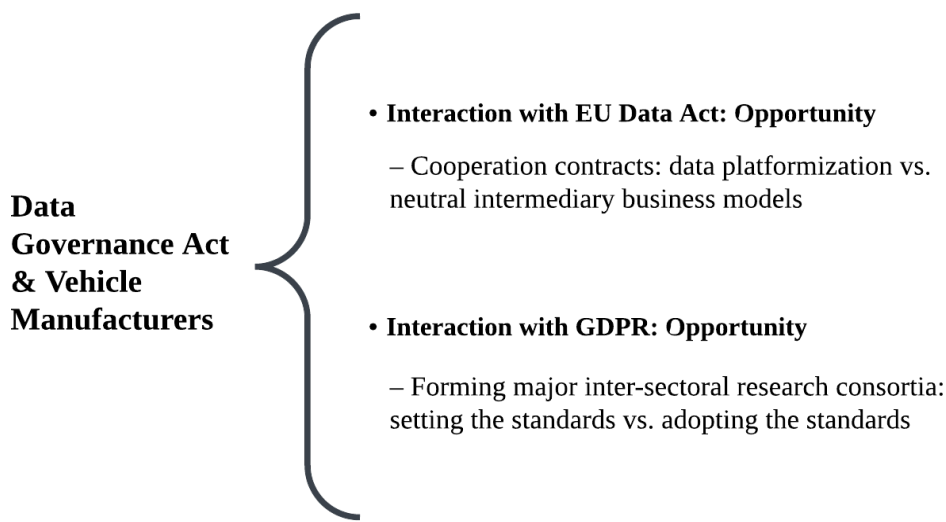

*Figure 5. The Interaction of EU Data Governance Act with Other Regulations*

Promoting big research in and across sectors and industries implies data normalization and enhanced data interoperability and portability capabilities. This is important specifically for private sector actors with large amounts of data, such as vehicle manufacturers. These actors, as discussed in the previous section, are subject to mandatory data sharing requirements under the Data Act. The Data Act requires private sector businesses to share data with their customers and their designated third parties in a readable and understandable way (data interoperability) and allow them to switch to a different service provider (data portability). Both data interoperability and portability require heavy technical investments by the businesses and are more difficult for the complex environment of cars with a myriad of physical and digital parts than purely digital contexts dominated by one central platform leader such as Apple's iOS (Huber et al., 2017: Hodapp and Hanelt 2022). Developing sector-specific interoperability standards usually takes more time than businesses can afford and even when standards emerge, businesses have found top-down standard definitions seldom work for everything (Kubicek et al. 2011), and that focusing on technical development does not necessarily lead to industry-wide adoption in the long-run (see Saadatmand et al. 2019).

---

[22] See, for instance, Section 8.4., Discriminatory bases for permission to process special category data: private sector research vs. public sector research

[23] See Data Act, Recital 2

Exchanging data with intermediaries for the sake of expanded research opportunities allows for the development of interoperability and portability standards with strong roots in practical scenarios. It allows complementors' to focus their attention on something tangible, and promotes engagement, incentive to dedicate resources, and collective identity, (Saadatmand et al. 2019). This, in turn, has an added value for more major businesses such as vehicle manufacturers. Promoting large-scale research opportunities allows the same standards to be created around the operations of vehicle manufacturers with larger datasets and allows these standards to be accepted by more trading partners. This affects community readiness across firm boundaries and thus achieving interoperability in dynamic value networks becomes easier. In other words, the network effect of the bigger firms affects the community readiness of other partners in the same ecosystem to naturally integrate with bigger firms (Zhao and Xia, 2014).

# 6. Digital Services Act

At first glance, the *Digital Services Act* is arguably the least applicable to vehicle manufacturers of the European data regulations. However, our analysis indicates that it may indeed matter to vehicle manufacturers. As its name conveys, the Digital Services Act aims at regulating the obligations of the intermediary platforms that provide digital service. Some paradigmatic examples of such service providers include, e-commerce marketplaces, video-sharing and media-sharing platforms, social networks, and collaborative economy platforms that act as intermediaries in their role of connecting consumers with goods, services, and content.

## 6.1. The Digital Services Act and E-Commerce Directive

Before the Digital Services Act, the *E-Commerce Directive*[24] set the foundational European legal framework for safe online services in the beginning of the millennium. The E-commerce Directive's aim has been to regulate information society platforms in the sphere of electronic commerce. It covers all service providing platforms such as those that are merely conduits of content (and have no or very little control over the content being exchanged), those that cache or store data, as well as platforms that host social or commercial transactions (e.g., publishing or distributing user content). According to the E-Commerce Directive a safe online environment would be provided by identifying, flagging, and removing illegal content. Illegal content is not explicitly defined in the E-Commerce Directive, rather it is left to the Member States to identify and take action against content that is defined as illegal because it is not compliant with an EU law or the law of a Member State. Some well-established and widely accepted examples of illegal content categories, according to the E-commerce Directive, include:

- Defamation
- Incitement to violence
- Child pornography
- Racist or xenophobic content
- Hate speech
- Terrorist content
- Commercial scams and frauds
- Breaches of intellectual property right

The E-Commerce Directive already establishes the foundational legal framework for online services in the EU and the Digital Services Act adopts and reincorporates the Directive extensively. The need for a Union-level regulation, however, emanates from two issues. First is the fragmented application of the Directive in Member States. As with any Directive, the E-Commerce Directive has left the enforcement of its requirements to the discretion of the Member States to set the necessary procedures and approaches. The EU Commission's impact assessment of the E-Commerce Directive (2020) argued that, with the exponential growth of the online intermediary service providers since the enactment of the E-Commerce Directive, both interpretations and implementation procedures needed to be harmonized.

Second, since the formulation of the E-Commerce Directive in 2000, the characteristics of online intermediary platforms have substantially changed. For instance, although online platforms did exist around the time of the E-Commerce Directive, their scale, reach and business models were in no way comparable to their current influence in the market and the functioning of our societies. An example is how, today many more economic activities such as transport and short-term rentals, besides the more common functions such as video sharing, music streaming, or online shopping, are performed through online platforms. Another example is the changes of online advertising that started with simple email lists and now happens based on targeted profiling of users through mixing data from multiple sources. Finally, while there are many different micro, small or medium size online platforms, the majority of users concentrate around a small number of very large online platforms. This turns the very large online platforms into de facto public and cross-border spaces for electronic commerce. These changes render the fragmented Member States approaches to creating safe online environments not only costly but also ineffective.

---

[24] Directive 2000/31/EC of the European Parliament and of the Council of 8 June 2000 on certain legal aspects of information society services, in particular electronic commerce, in the Internal Market ('Directive on electronic commerce').

The Digital Services Act, as a regulation, harmonizes these fragmented approaches and focuses on creating certainty in terms of what the online intermediary service providers need to do to avoid liability. A list of the Digital Services Act's amended points is extended in 5.3. Here, we touch upon one amended issue, the "liability regime" (footnote) that arguably reveals the logic of "liability" in the sphere of digital services according to the EU's most iconic regulations on digital services. This is important since many vehicle manufacturers are connected to digital services (e.g., cloud services, telematics functions and applications provided through online intermediaries). This connection poses reputational risks to vehicle manufacturers, as argued in 5.5.

The E-commerce Directive does not specifically spell out the conditions under which the intermediaries are deemed liable. It does however identify explicitly the conditions under which an intermediary can be *exempt* from liability. These two conditions include: a) having no knowledge of the illegal content, and b) acting expeditiously to remove or to disable access to the information upon obtaining information about such content.

It must be mentioned that for the provider to prove having no knowledge (first condition for liability exemption), they must prove that they neither had "actual" knowledge nor were they aware of the facts and circumstances from which an illegal activity or information is apparent (this type of knowledge is referred to as constructive knowledge). For instance, actual knowledge, as its name suggests, refers to when the intermediary actually knows or is informed about the existence of illegal content. Constructive knowledge however is a more implicit form of knowledge. An example is when, given the criteria used in a service provider's algorithmic indexing, that service provider should know how those criteria will affect the presentation of content on the platform.

The EU Commission's Impact Assessment of the E-Commerce Directive[25], identifies three tensions in relation to different adaptations of the liability problem in the E-commerce Directive:

- **Active role by means of constructive knowledge:** According to the EU Commission's Impact Assessment (2020), however, various courts across Member States have interpreted this condition differently over the years. One example is that some courts have not taken notice of the fact that today, automatic, algorithmic ordering, displaying, and tagging or indexing of the content are necessary to make content findable in the first place. Thus, any intermediary providing such content could presumably be knowledgeable about this dynamic. That is, they could be considered to have constructive knowledge about the fact and circumstances from which illegal content can ensue, and therefore to have had an "active role" in the illegal content dissemination.

- **Active role by appropriation impression**: Another example is that some courts across different Member States have interpreted the intermediary's role as "active" in disseminating illegal content if the intermediary has presented the content in a manner that a reasonably informed user could conclude that the platform is the author or responsible for such content. In other words, the presentation of the content creates an "appropriation" effect. However, not all courts across different Member states have adopted this view in a unified way.

- **The Good Samaritan paradox[26]:** Finally, in some cases, the proactive measures of intermediaries to detect illegal activities (even by automatic means) has had the potential to be interpreted as "having knowledge" of or "control over" the illegal content and by extension assigning an "active role" to them in disseminating illegal content. This legal uncertainty has created contradictory incentives for service providers to engage in voluntary measures for detecting, removing, or disabling access to, illegal content.

---

[25] Commission Staff Working Document Impact Assessment Accompanying The Document Proposal For A Regulation Of The European Parliament And Of The Council on a Single Market For Digital Services (Digital Services Act) and amending Directive 2000/31/EC, Brussels, 15.12.2020 SWD(2020) 348 final.

[26] The Good Samaritan paradox is discussed widely in relation to a corresponding piece of legislation in the United States that sets up the conditions for exemption of service providers from liability. These conditions are stated under Section 230 of Title 47 of the United States Code that was enacted as part of the Communications Decency Act of 1996, and they have led to similar misinterpretations and complexities.

The Digital Services Act amends these fragmentations and uncertainties in two ways. First, it adds one more section to the conditions for liability exemption of Article 14 in the E-Commerce Directive. The liability exemption conditions in the Digital Services Act are presented under Article 6 which adopts both conditions in Article 14 of the E-Commerce directive and adds that liability exemption does not apply if:

> *"The online platform presents the specific item of information or otherwise enables the specific transaction at issue in a way that would lead an average consumer to believe that the information, or the product or service that is the object of the transaction, is provided either by the online platform itself or by a recipient of the service who is acting under its authority or control".*

Thus, per Digital Services Act, "presentation in the form of appropriation" can arguably imply having an "active role" in the illegal activity and therefore eliminates an actor's chance to be exempt from liability. Further, Recital 26 of the Digital Services Act explicitly provides the legal certainty that activities carried out in good faith and in a diligent manner to detect and remove illegal content will not render unavailable the exemptions from liability set out in Article 6 of the Digital Services Act.

### 6.2. Summary of the Digital Services Act

Similar to the GDPR, the Digital Services Act aims at protecting users' fundamental rights, but its focus is on promoting safe online environments. Online environments are safeguarded under the Digital Services Act by preventing risks such as disinformation or election manipulation, fake news, cyber violence, online harms to minors, targeted advertising by profiling minors or based on special categories of personal data (e.g., ethnicity, political views or sexual orientation), and the use of dark patterns to manipulate users into choices they do not intend to make (e.g., buying certain items online). In general, the Digital Services Act aims at protecting users' rights and the EU's internal market by:

- Protecting economic rights of platform users

- Protecting platform users' fundamental rights such as the freedom of expression and information

Two major problems have incentivized the European Union to frame a Union-level legislative instrument to protect EU citizens from harmful content and commercial strategies in online environments. First, in the absence of a unified approach, each Member State across the EU has proceeded with devising their own regulatory instruments to control the sphere of online services. This has resulted in extreme regulatory fragmentation in Europe affecting SMEs' innovation and outreach in the European market as they do not have enough resources to comply with different Member State requirements. It has also affected innovative investments in Europe by tech companies in the U.S and China due to the increased compliance costs that the fragmentation will cause across the European Member States. Second, some key concepts that are crucial in interpreting harmful content/behavior and therefore liability/compliance requirements needed to be updated to match the technological advancements that bestow different capabilities on providers of online services.

| Digital Service Act (DSA) | |
|---|---|
| **Problem** | **Instrument** |
| **Legal** | |
| • Extreme legal fragmentation across different Member States | • Devising a legislative instrument that **unifies the goals and approaches** of achieving them on the **Union level.** |
| • **Outdated concepts** and categories that do not match the consequences of technological advancements constituting the backbone of digital services and platforms | • **Clarifications** for new types of services in the Internet stack **not clearly fitting** in the categories of the **E-commerce Directive**.<br>• **Clarification** of where a service **cannot** benefit from the **liability exemption**. |
| ● **Lack of clarity** as to when a service provider has done their **due diligence** to address the risks brought by the dissemination of illegal content, goods or services online | ● Clarifying the necessary **obligations** that define "**due diligenc**e". |
| ● **Lack of procedural clarity in supervisory and enforcement obligations** | Clearly stating the procedures to prove compliance with supervisory and enforcement requirements including:<br>● Use of trusted flaggers<br>● Digital Service coordinators tasks<br>● Notice and action compliance requirements:<br>● Regulating at the country of use |
| <div align="center">**Technical**</div> | |
| ● Lack of self-auditing approaches | ● **Mandating** approaches for **risk auditing** and **risk mitigation** for **certain categories** of online service providers |
| ● Leck of technical knowledge and means for supervising online intermediaries | ● **Mandating the use of** trusted **flaggers** that have the technical capability to flag risk |
| <div align="center">**Economic**</div> | |
| ● **Negative** effects on **innovative capabilities** of service providers due to the obligations to abide by **multiple fragmented supervisory rules** across different member States | ● Providing a **unified approach** at the **Union** level |
| ● The economic burden on the **SMEs** to be compliant with over 27 Member States' regulation and devise **appropriate auditing or reaction methods** | ● Devising a **tiered approach** to liability with **SMEs** facing **minimum** compliance **requirements** |
| ● **Low innovation investment** from **international service** providers from the U.S. or China in important areas such as mobility or food due to **increasing costs of fragmented compliance requirements** | ● Providing a **unified approach** at the Union level |

*Table 3. Summary of the EU Digital Services Act (DSA)*

The Digital Services Act aims to mitigate these problems and achieve safe online environments by:

#### 6.2.1. Updating Key Concepts and Requirements

The Digital Services Act does not substitute the E-commerce Directive but adopts substantial parts of it. It does however further clarify several issues described below:

- **Defining categories of online intermediary service providers:** the Digital Services Act sets out the requirements that cover "intermediary services". The Digital Services Act adopts the E-Commerce Directive definition of "intermediary services" and uses it as a catch-all term that includes providers of *all* online services from more backend services – such as providers of core internet infrastructure, internet access, domain name systems operators supported by other types of technical services such as content delivery networks, or cloud infrastructure services, caching services, hosting service providers (e.g., cloud services, web hosting) – to more front-end service providers such as online marketplaces, app stores, collaborative economy platforms, social media platforms, and online search engines. However, the Digital Services Act adds a category called "very large online platforms/search engines" and distinguishes it from other online platforms.
- **Definition and categories of illegal content:** The Digital Services Act explicitly defines illegal content as any information that is not compliant with EU law or the law of a Member State" (Article 3). The E-commerce Directive also implied this definition but did not explicitly state it. The Digital Services Act also adds some more categories to the prohibited types of illegal content (e.g., dark patterns in Article 25 and personal data profiling in Article 26).

- **Supervisory and enforcement powers:**
  - **Trusted Flaggers:** The E-Commerce Directive recommended that platforms should cooperate with "trusted flaggers" who assist them in identifying illegal or harmful content. This could be seen as a system for platforms to outsource some parts of their responsibility for content moderation. This was however recommended as a voluntary system. The Digital Services Act mandates the use of trusted flaggers and introduces harmonized criteria for actors to be designed as trusted flaggers. These criteria include (Article 22):
    - it has particular expertise and competence for the purposes of detecting, identifying and notifying illegal content,
    - it is independent from any provider of online platforms,
    - it carries out its activities for the purposes of submitting notices diligently, accurately and objectively.

    Trusted flaggers must publish easily understandable and detailed annual reports about the notices submitted that include information on the types of illegal content reported, and the actions taken by the online platforms.

  - **Digital Service coordinators:** help the European Commission to monitor and enforce obligations in the Digital Services Act. They are the first point of contact for enforcing the Digital Services Act and have responsibilities such as coordinating national authorities to ensure cooperation and enforcement of the Digital Services Act, monitoring and enforcing their Member States' laws on platforms, imposing fines, and authorizing trusted flaggers. Digital Service coordinators were not directly regulated before the Digital Services Act.

  - **Notice and action compliance requirements:** before the Digital Services Act, the E-Commerce Directive regulated the obligation for notice and removal of illegal content, but it did not set any procedural obligations for how to comply with this requirement. The approach was left to be decided by Member States. The Digital Services Act clarifies and harmonizes the procedure to comply with this requirement in Article 16.

  - **Regulating at the country of use:** According to the e-Commerce Directive the online intermediary platforms were legally obligated to the laws of the country where they were established and not where they were used. The Digital Services Act removes the "country of Origin" rule and regulates online intermediary platforms across Member States where they are used.

### 6.2.2. Devising a Tiered Approach to Confirming Due Diligence and Liability

The Digital Services Act recognizes 4 levels of due diligence corresponding to *four* levels of control over the content by the online intermediaries. Higher levels build upon lower levels. For instance, the control that a merely caching or storing platform has over the content is lower than the control that an information distributing platform has over its content. The lower the control, the lower the level of requirements to demonstrate the platform's "due diligence" for protecting users' rights and safety.

The first level sets the bare minimum requirements of "due diligence" and applies to all intermediary services mentioned above. The strictest requirements (Level Four) are naturally imposed on very large online platforms (VLOPs) and very large online search engines (VLOSEs) who also must comply with all the previous levels of due diligence summarized in the table below. The highest level focuses on those organizations that not only have significant control over caching, storing, and distributing large amounts of content, but also can cause more significant damage due to their extensive outreach and network of users.

| Levels of control/liability | Proof of Due Diligence Requirements |
|---|---|
| **Level One:**<br>**All Intermediaries** | • All intermediary services, that **host**, **cache, or store information**, must designate and make publicly available a single point of contact for the public authorities and users to communicate issues related to Digital Services Act with.<br>• Designate a representative in the EU if they lack establishments in the EU.<br>• Define user-friendly and clear terms and conditions explaining their methods of content moderation (including the use of automated technologies).<br>• Must publish annual **transparency** reports containing information on their content moderation practices (not applicable to SMEs). |
| **Level Two:**<br>**All Hosting Services** | • All hosting services including online platforms that **store data** must utilize a "notice and takedown" mechanism that allows users of the service to report allegedly illegal content.<br>• Must promptly review the notices and decide whether to remove the content and notify the user with a statement of reason explaining the action. |
| **Level Three:**<br>**Online Platforms**<br><br>(Does not apply to SMEs) | • All online platforms that **not only store but distribute user info** must allow users to complain about a decision made by the platform, either to the platform's internal complaint-handling mechanism or an independent out-of-court dispute settlement mechanism that can make a non-binding decision on the content or account in question.<br>• When handling notices on allegedly illegal content, online platforms must give priority to notices received from so-called "trusted flaggers", i.e., associations certified to make accurate notices.<br>• Must design platforms without dark patterns deceiving, manipulating, or otherwise misleading users from making free and informed decisions.<br>• Advertisements should be clearly marked as such and contain information on the advertiser.<br>• Are prohibited from using personal data of minors or special category data to present targeted advertisements.<br>• All **marketplace** platforms must verify the identity of traders who use their platform to conclude distance contracts with consumers. |
| **Level Four:**<br>**VLOPs & VLOSEs** | • Must appoint a compliance officer.<br>• Must provide data access to supervisory authorities.<br>• Must produce **risk** reports.<br>• Must have independent auditors verify their compliance with the Digital Services Act.<br>• The European Union has a supervisory authority over VLOPs and VLOSEs. |

*Table 4. The EU Digital Services Act's Tiered Approach to Due Diligence Requirements*

### 6.3. The Digital Services Act and GDPR

- **Design Transparency requirements - vehicle manufacturers vs. online platforms:** When it comes to the automotive industry, the Digital Services Act should apply mostly to mobility service providers (e.g., Uber, leasing, predictive maintenance services), rather than original vehicle manufacturers. However, paying attention to the reason the Digital Services Act was devised reveals the complex connection vehicle manufacturers can develop with this Act. Regulating intermediaries of digital services is nothing new. In fact, most of the protective measures adopted by the Digital Services Act are a confirmation of similar measures already covered under the E-Commerce Directive which has been in effect for over two decades. What is new is the type of intermediary digital services in scope. When the E-Commerce Directive became effective at the end of the 1990s, some online platforms did exist. However, their scale, reach and business models were incomparable to their current influence in the functioning of the market and society.

   This evolution is ignored in, for instance, the different ways the E-Commerce Directive has been (mis)interpreted and enforced across different Member States when it comes to defining an "active" role for digital service providers in controlling the content and therefore market.[27] An active role has been equated to having knowledge or direct control over the customer-facing services to the extent that a reasonably informed user tends to conclude that the platform is the author or responsible for such content or service. Anything that fell out of this conclusion has been oftentimes assigned a "passive" role and therefore has not been subject to (enough) liability.

---

[27] Impact Assessment Accompanying the document Proposal For A Regulation Of The European Parliament And Of The Council on a Single Market for Digital Services (Digital Services Act) and amending Directive 2000/31/EC, Brussels, 15.12.2020 SWD (2020) 348 final.

This interpretation is problematic since the technological advancements in algorithmic ordering, displaying, and tagging or indexing of the content that a service provider stores, play a crucial role today in making any content findable or accessible (e.g., L'Oréal vs eBay court case). This arms-length control power over what the customers get exposed to implies an active role, according to the Digital Services Act.[28]

Some very large online platforms (sometimes referred to as VLOPs) and cloud service providers such as Google and Apple are already providers of apps and telematics services to prominent car manufacturers. Telematics service providers specialize in collecting, processing, and managing connected car data. They often work with car manufacturers and offer services like remote diagnostics, vehicle tracking, and usage-based insurance. They are also the most influential app developers that develop mobile applications or software platforms that interact with the car's systems. These apps may provide services like navigation, entertainment, or real-time vehicle monitoring. The range of data collected and processed in these two areas reveals substantial amounts of information about the users to these VLOPs allowing them to create user profiles based on data categories prohibited in the Digital Services Act for targeted advertising. Yet, in the disturbing and controversial reports about car brands' privacy breach (e.g., Mozilla report on vehicle makers' excessive breach of privacy requirements), the public rarely separates the vehicle manufacturer from the platform and service providers. This phenomenon is understood in terms of a social contract for securing privacy and trust – in which individuals within a given context discriminately share information with a particular set of (unwritten) obligations in mind as to who has access to the information and how it will be used, rather than a written privacy contract (Martin 2018). Within the view of privacy and trust as a social contract, vehicle users have trusted the vehicle brand they have known for years and trusted that they will be protected according to appropriate norms of privacy.

Similarly, in understanding the active role of players in breaching privacy or fundamental rights, the scholarship on privacy is emphasizing that the goals and obligations of an online platform (who create a platform for other actors to interact) is essentially different from that of an original manufacturer (who makes and sells products and services directly to consumers). Whereas a manufacturing firm gains market share through product quality and pricing, the platform maintains market dominance through its governing policies (Martin 2023). A vehicle manufacturer's market power is related to the demand for the product and its brand. But the product and its brand may not be directly related to the amount of data the firm holds. That is, a manufacturer might hold a large amount of data, yet have a low market power. As the Lead Enterprise Architect in a truck manufacturer mentioned:

*"We are not sitting here on a daily basis trying to figure out how to make money with others' data, we have a business and that's to make and sell trucks"*.

On the contrary, for an online platform, it is often the manipulation of data governance policies (e.g., to rank their own services and products higher or to create user profiles for targeted advertising) that can bring about the most lucrative market segments (Martin 2023). Thus, a platform may be relatively less transparent about the context of exchange and purpose of data collections and processing, whereas a vehicle manufacturer is more likely to collect and use data for the purposes it directly states and therefore is benefitting from a higher level of trust by users. If we adopt the perspective of privacy that considers individuals as engaged in controlling their privacy dynamically and based on the contextual conditions (see Acquisti 2023) and social contracts, then a privacy or fundamental rights breach bears heavy reputational implications for the manufacturers who integrate online platforms and services under their trusted brand.

At the same time, the Digital services Act's flagging, complaint filing, annual risk and transparency reports as well as auditing requirements are intended to make it increasingly easier for both users and trusted flaggers to identify and report risks associated with the conducts of these platforms. Thus, although the Digital Services Act does not directly address vehicle manufacturers, it poses a silent liability risk. One area that gives prominence to the threat of this silent liability is the fact that the Digital Services Act provides clear conditions for liability exemption on the Union level, but not for conditions under which liability is incurred. Instead, these conditions are determined by other EU and national laws (Turillazzi et al. 2023). Thus, room for interpreting how "active" the role of each ecosystem actor is, will be much more open. It is, for instance, possible that the active role of the manufacturer is evaluated against the principle of "privacy by default and design" under Article 25 of the GDPR. This principle might imply that manufacturers should be transparent about how the design of the different systems in the car (e.g., vehicle platform or telematics systems) affect the collection and processing of data subjects both to the users and the GDPR data protection authorities (DPAs).

---

[28] Paragraph 112, Commission Staff Working Document Impact Assessment Accompanying the document Proposal For A Regulation Of The European Parliament And Of The Council on a Single Market for Digital Services (Digital Services Act) and amending Directive 2000/31/EC, , Brussels, 15.12.2020 SWD(2020) 348 final.

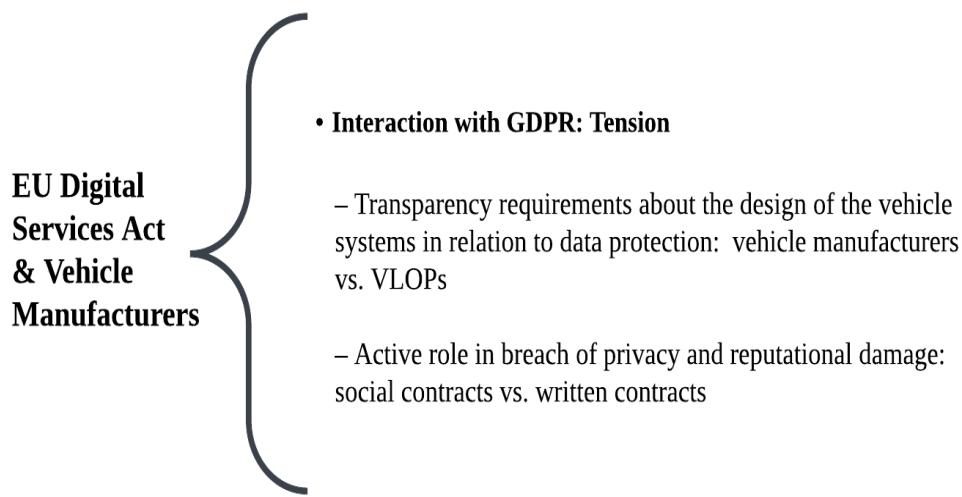

*Figure 3. The Interaction of EU Digital Services Act with Other Regulations*

Another argument is that based on article 14 of the Digital Services Act, if liability exemption does not apply when the information is presented in a way that gives the "appropriation" impression. That is, when the online platform enables the specific transaction at issue in a way that would lead an average consumer to believe that the product or service is provided either by the online platform itself or by a recipient of the service who is acting under its authority or control. Although this section directly addresses online platforms that host various types of transactions, it can imply car manufacturers since upon buying or leasing or using a vehicle, the customer usually presumes that they are facing a product under the vehicle manufacturer's brand and not under Google's or Apple's CarPlay. What matters is the slight but important changes in the logic and underlying assumptions that lead us to believe a party should or should not be held liable to a breach of safety, privacy, or trust. Therefore, although vehicle manufacturers are safe from scrutiny at first glance, they can be implicated as partners and facilitators of the breach.

# 7. Digital Markets Act

The Digital Markets Act is one of the EU's major legislative instruments for regulating competition. Whereas the Digital Services Act's focus is more on protection and safety in relation to some dominant actors (e.g., very large online platforms), the Digital Markets Act aims at regulating commercial competitiveness in relation to a specific group of dominant market players (referred to as gatekeepers).

## 7.1. Digital Markets Act and the Treaty on the Functioning of the European Union

The Digital Markets Act amends some of what is perceived to be structural shortcomings in the already existing competition laws such as the Treaty on the Functioning of the European Union[29] (TFEU, 1950s). Some of the main focuses of the Digital Markets Act are already addressed by the TFEU. Examples include regulatory mandates to prevent anti-competitive agreements and concerted practices between companies (Article 101), or to prevent a company from abusing its dominant position. This includes, for instance, directly or indirectly imposing unfair purchase or selling prices, limiting production, markets or technical development to the prejudice of consumers, as well as applying dissimilar conditions to equivalent transactions (Article 102). However, the TFEU's regulatory instruments apply to all the operations of all players in a market equally. This is while, policy makers argue that since the treaty's inception, a certain type of market player has emerged, associated with digital platform ecosystems, and this has led to new market relations and dynamics. The distinct characteristics of these players and their operations dynamics has inspired the EU to refine and create new concepts, doctrines, and methodologies, and thereby regulatory instruments that address such distinctions and changes.

For instance, the growing availability of algorithm-based technological solutions facilitates the monitoring of competitors' conduct and creates increased market transparency. This allows increasingly concentrated markets to create incentives to compete less vigorously without any direct coordination (the so-called tacit collusion).[30] In their final report on proposing a "Competition Policy for the Digital Era", the EU commission touches upon a few concepts and market dynamics that need to be reviewed.[31] These core concepts and issues exemplify the changed market dynamics and relations based on which the EU Commission argues for new competition laws. Some examples of the changing market dynamics, according to the EU Commission, include:

- **Market definition:** in the case of multisided platforms, the interdependence of the sides of a platform becomes a crucial focus of analyzing the market, whereas the traditional role of market definition has addressed different market players in isolation.

- **Measuring market power:** the data-related market dynamics make it difficult to measure market power. For instance, in the case of dominant market players such as digital platforms, even in an apparently fragmented marketplace, there can be market power. This kind of market power is linked to the concept of "unavoidable trading partner" or what has sometimes been called "intermediation power" in the area of platforms (p. 4). In the data-related markets, even non-dominant marker players can turn into unavoidable trading partners. For instance, if data that is not available to market entrants provides a strong competitive advantage, its possession even by a non-dominant company may lead to their market dominance (see also the European Commission's New Competition Tool 2021).

- **Competition law and regulation:** the new proposals are being developed to complement rather than substitute for previous regulations. Previous competition laws were designed to react to ever-changing markets and provide a framework for refining some of the concepts and methods of regulating markets as new dynamics and relations emerge.

---

[29] -Consolidated version of the Treaty on the Functioning of the European Union, OJ C 326, 26.10.2012.

[30] Proposal for a Regulation by the Council and the European Parliament introducing a new competition tool, Ref. Ares (2020)2877634 - 04/06/2020.

[31] Publications Office of the European Union final report on Competition Policy for the Digital Era, 2019

- **The consumer welfare standard:** Businesses are affected by practices of platforms at a different speed and scale. This calls for rethinking both the timeframe and the standard of proof in light of the likelihood of cost of error. For instance, in assessing harm to consumers, the focus has historically been on the short-run impact of pricing. Preexisting antitrust regulations have mainly calculated costs based on the probability of an error occurring, which were more predictable with industrial production. With digital platform organizations, however, the scale, timeframe, and opacity of harms is different, and errors often have long-lasting effects due to a platform's sticky market power. Therefore, the standards based on which consumer welfare should be evaluated need to change (e.g., fair prices vs. fair innovation opportunities).

- **The error cost framework:** Pursuant to changing standards to assess harm, the frameworks based on which we assign costs and encourage or prohibit actions should also be reviewed. For instance, as argued before, in the case of digital platforms, the effect of wrongful actions might be obvious only in the long run while at the same time those errors can have a long-lasting effect on consumers. Thus, it might not be easy to directly test and determine the type and scale of the costs of a particular action. Therefore, previous error cost frameworks that were based on probability of wrongful action might not be relevant in the case of digital platforms and ecosystems. Instead, the policy makers are arguing that "potential harm" (and not probability of harm) should be the sufficient basis of antitrust regulations to condemn a platform's actions. Similarly, the burden of proving harmful action should not be on the consumers as it was previously the case[32]. This is because it would be increasingly difficult for them to track and determine the cost of harm that might happen in the long run. Instead, the burden of proof should be on platforms to argue for how beneficial and pro-competition a particular action is. With this background, the DMA amends the antitrust laws and reverses the burden of proof on the gatekeeper to prove that their conduct is legally compliant[33].

On a general level, in their proposal for a Competition Policy for the Digital Era, the EU Commission focuses on three main points that delineate the particular characteristics of the digital era in which digital ecosystems and platforms play consequential roles in the market. These characteristic points include:

1. **Extreme returns to scale:** cost of production of digital services is much less than proportional to the number of customers served.

2. **Network externalities:** to convince users, it is not enough for a new entrant to offer better quality and/or a lower price than the incumbent does. Due to the lack of multi-homing, interoperability and portability, the users would be reluctant to change service providers.'

3. **The role of data:** The evolution of technology has made it possible for companies to collect, store, and use large amounts of data necessary to develop new, innovative services and products. Data is thus a major competitive parameter whose relevance will continue to increase. Digital platforms providing a variety of services (e.g., Meta offers social media, transaction, messaging services) hold disproportionately large amounts of data taken from multiple points.

Below, we explain how these changing concepts and relations arguably separate digital ecosystems and platforms from other incumbent and often dominant market players and hence call for a specific legislative focus.

### 7.2. Summary of the Digital Markets Act

The Digital Markets Act is applicable only to a select number of platforms that are designated as gatekeepers by the EU Commission and parliament. On September 6th, 2023, the European Commission identified six gatekeepers, Alphabet, Amazon, Apple, ByteDance, Meta, and Microsoft. The EU identifies these gatekeepers based on several characteristics such as extreme scale economies, very strong network effects, a significant degree of dependence of both business users and end users, lock-in effects, a lack of multi-homing for the same purpose by end users, vertical integration, and data driven-advantages. The Digital

---

[32] Article 2 of Council Regulation (EC) No 1/2003 of 16 December 2002 on the implementation of the rules on competition laid down in Articles [101 and 102 TFEU]

[33] For comparison, see Article 2 of Regulation No 1/2003 and Article 6 of the DMA.

Markets Act presents three main bases that arguably grant an enterprise with a gatekeeping position that substantially affect market competition and therefore should be subject to the Act's strict obligations. These three bases correspond to the three main points that delineate the particular characteristics of digital ecosystems and platforms mentioned in the EC's proposal for "Competition Policy for the Digital Era", and include:

- **Extreme scale economy**: In the extreme scale economic model, there are nearly zero marginal costs to add and connect end users, businesses, and individuals from a myriad of heterogeneous areas (the ubiquitous effect). Then, by acquiring distinctive, unique complementary activities under their direct control, large platforms own various stages of the production process within the same industry and thereby tie the users so strongly to their processes and transactions (the vertical integration effect) that, in practice, switching service providers becomes possible only with considerable effort and expense (the lock-in effect).

- **Mega networks:** these gatekeepers all provide core platform services, including online cloud computing, web browsers, online search engines, online intermediation, online social networking, the whole online advertising supply chain (networks, exchanges, and intermediation), video sharing, and mass interpersonal communication. A gatekeeper enjoys, on average, a minimum of 45 million monthly end users with an annual turnover of at least €7.5 billion within the EU in each of the past three financial years or an average market valuation of at least €75 billion.

- **Data driven advantages**: Now that a vast scope of users is strongly dependent on one player's services, they leverage the extensive effects of the data yielded through the use of their services, and they have a proprietary hold on the data. This designates that player as a *gatekeeper*.

The Digital Markets Act aims to regulate these gatekeepers with the following goals:

- Increase competitive capability of smaller businesses in relation to undertakings with considerable economic power.

- Offer fair market competition and balance the bargaining power of existing or new market operators based on innovation and efficiency.

- Offer competitive prices for services.

- Unify the divergent regulatory solutions that were proposed or adopted on a national level to address unfair market advantages of gatekeepers.

- Prevent gatekeepers from adopting different conditions in different EU Member States (as some of them have, and thereby creating multiple, disparate, and unfair market conditions for users of the core platform services).

- Prevent Member States from applying national rules to market competition and thereby create fragmentation within the EU's internal market.

Although major vehicle manufacturers hold a network of scale, they do not enjoy extreme-scale economies with almost zero costs for reaching their network of users. On the contrary, the cost of manufacturing is enormous. In other words, they are not gatekeepers despite their mega user network. However, the Digital Markets Act potentially implicates direct effects on vehicle manufacturers on several fronts once we remember that some gatekeepers are the major providers of cloud-based functions, telematics applications, and even fully featured operating systems (e.g., Android Automotive) to the vehicle manufacturers.

| Digital Markets Act (DMA) | |
|---|---|
| **Problem** | **Instrument** |
| **Legal** | |
| • Lack of market power for smaller businesses or end users to file complaints against the gatekeepers' unfair practices | • Provide the whistleblowing system to alert authorities against gatekeepers' unfair practices |
| • Fragmentation in national level legislations | • Designate European Commission as the sole enforcer of these regulations to prevent fragmentation in legislative ruling and practice in Member States |
| **Technical** | |
| • Vertical integration and exclusive control over various stages of product/service development | • Define **obligatory technical requirements** to demand gatekeepers to:<br>– Offer greater choice, such as the choice of certain software on a user's operating system.<br>– Ensure, free of charge, effective interoperability with, their operating system, hardware or software features available in their own complementary and supporting services and hardware. As well as interoperability with their number-independent basic functions<br>– Ensure that unsubscribing from core platform services is as easy as subscribing (portability).<br>– Give business users access to their marketing or advertising performance data on the platform. |
| **Economic** | |
| • Larger market shares overweighing innovative capabilities of smaller players | • Sets out rules for defining large online platforms designated as important gatekeepers with economies of scale and controlling powers |
| • Perceived unfair business practices and prices by large online platforms designated as important gatekeepers | • Define **prohibiting rules** to prevent gatekeepers from imposing unfair conditions on their service users, including prohibiting gatekeepers to:<br>– restrict business users of platforms.<br>– rank their own products or services higher than those of other companies.<br>– prevent developers from using third-party payment platforms for app sales.<br>– pre-install certain software applications or prevent users from easily un-installing them. |
| • **Fragmented market conditions** created by gatekeepers in different EU member States | • Providing a set of **unified requirements** at the Union level that requires big tech companies to treat organizations across the EU equally. |

*Table 5. Summary of the EU Digital Markets Act (DMA)*

### 7.3. The Digital Markets Act and Data Act

- **Removed lock-in effects and increased interoperability/portability capabilities for vehicle manufacturers vs. gatekeepers:** first, the Digital Markets Act potentially removes the costly lock-in commitments to software or cloud services of one provider. When the lock-in effects are abated, the vehicle manufacturers have more freedom to switch between cloud functions and services. It must be considered that many of these core platform services are offered as a response to market problems and failures (Jacobides et al. 2024). Google, Apple, and Microsoft, for instance, specialize in telematics services and in-vehicle apps and thus offer vehicle manufacturers with indispensable functionalities and network values. The interorganizational value architectures between manufacturers and core platforms are based on modular technologies that facilitate production, integration, and further innovation (Cennamo et al. 2023). Therefore, we do not assume the role of the gatekeepers to be that of the complete winners that "take it all" (see Jacobides et al. 2024). The strategic opportunity for vehicle manufacturers involves the possibility to reconsider the "imbalance of rights and obligations" (Cremer et al. 2023) with organizations deemed gatekeepers. This enables vehicle manufacturers to avoid becoming locked into particular vendors, essentially enabling their vendor relationships to be more like real options (McGrath 2004) rather than indirect economic externalities (Jacobides et al. 2024) that cannot be accounted for in the transaction terms between the manufacturers and core platform providers. The focus is thus on the potential opportunity that the Digital Markets Act is extending to reconsider and redefine contractual fairness.

For instance, some vehicle manufacturers have rejected Google Automotive Services (e.g., Google Maps) which collects battery information, vehicle's GPS location, navigation details. These manufacturers believe that the data collected through the use of these applications – together with the data collected through the fully featured operating systems (such as Android Automotive that collects data from sensors) — endow the tech industry with a greater control over the automotive world than ever. In fact, many believe that their market dominance and the ensuing abuse of contractual practice, is mostly fueled by information collection and control rather than market share (Marty and Warin 2020). Since the Digital Markets Act is imposing requirements, such as offering a greater choice of services on operating systems, and free-of-charge interoperability on the side of gatekeepers, vehicle manufacturers can realign their position and eventually the ecosystem architecture, i.e., rules of the game (Jacobides et al. 2024). These regulatory instruments in the Digital Markets Act are intended to reduce the governance rigidity resulting from inseparability of manufacturers and platform providers, increase the number and heterogeneity of complementors (e.g., map and navigation service providers), and increase the complexity of the bargaining processes of different sides in the ecosystem (Uzunca et al. 2022). Under these circumstances, manufacturers can increasingly opt for a mix of different services and systems. This diversity of options possibly opens ways to negotiate new data exchange contracts or knowledge exchange agreements.

It must be mentioned that the Digital Markets Act alone lacks sector-specific regulatory instruments and therefore cannot regulate the intended contestability and fairness seamlessly. However, the overlap and complementarity between various regulations could provide the foundations for successful outcomes (see Larouche and de Streel, 2021). For instance, on the one hand, the Digital Markets Act imposes mandatory requirements on gatekeepers to offer greater choice of software and applications on users' (manufacturers') operating systems, as well as ensuring effective and free-of-charge interoperability between hardware and software features. On the other hand, the Data Act, allows manufacturers — as the users of gatekeepers' functions and applications — to request access to the data generated through these apps and functions. Compatibility of services and applications and the core platform could be an issue if this was a stand-alone legislation. It is not, however. It is a wave set of legislations and the compatibility issues in changing service providers are covered under the interoperability and portability instruments in both the Digital Markets Act and Data Act.

Additionally, by removing lock-in effects, the manufacturer's willingness to share data produced through cloud-based functions and telematics applications with their own users will be less encumbered by the gatekeeper's dominating power. The Digital Markets Act mandates increased interoperability and portability capabilities upon gatekeepers. This means increased interoperable and portable data for vehicle manufacturers, who in turn, can share that data with users in a more standardized format. Thus, the Digital Markets Act will potentially have easing effects on complying with the Data Act as well.

### 7.4. The Digital Markets Act, GDPR, and Data Act

- **Unilateral contractual advantage for vehicle manufacturers vs. gatekeepers:** finally, it is not just the business-to-business advantages in the interaction between the Digital Markets Act and the Data Act that can benefit vehicle manufacturers. Vehicle manufacturers can initiate new business models or contractual grounds (with their users by turning into the user's designated third party in data sharing. For instance, under the Data Act, the vehicle manufacturer can easily become the designated third party that a user of a gatekeeper's application would like to share that application's data with. The Data Act (Article 5(3)), however, prohibits gatekeepers from requesting for or being granted access to users' data generated by the use of a connected product or related service or by a virtual assistant. The Data Act (Recital 40) also prohibits gatekeepers from combining certain data from different sources without consent. It also imposes on gatekeepers an obligation to ensure effective rights to data portability for their users under Article 20 of the GDPR. At the same time, the Digital Markets Act (Article 5(2)-b-c) prohibits gatekeepers from combining or cross-using personal data from the relevant core platform service with personal data from any further core platform services or from any other services provided by the gatekeeper or with personal data from third-party services. Thus, vehicle users who engage with gatekeeper applications and systems in the vehicle, can restrict gatekeepers' access to their personal data recorded through the use of vehicles through contractual terms and conditions.

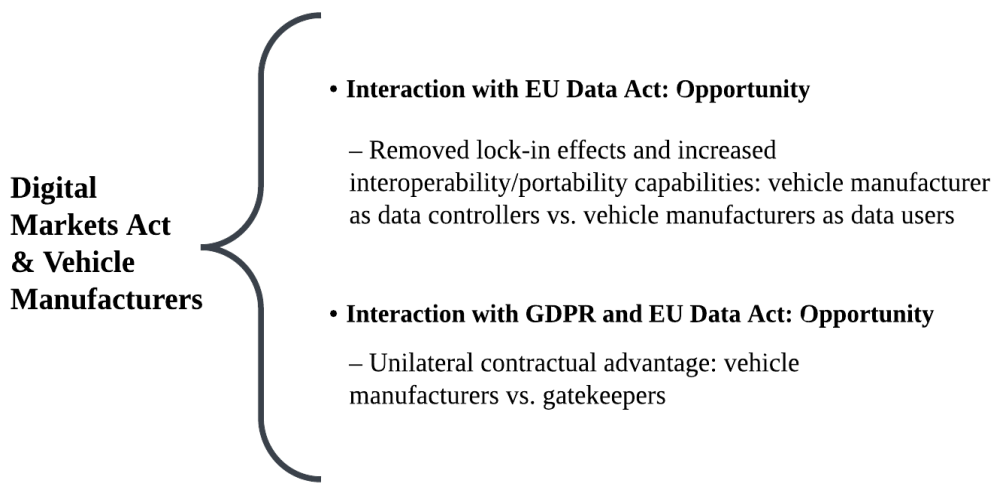

*Figure 4. The Interaction of the Digital Markets Act with Other Regulations*

The combination of these regulatory instruments (GDPR, Data Act, Digital Markets Act) bestows upon manufacturers an almost unilateral advantage. They can be designated third parties with whom the vehicle users would like to share both their personal and IoT data generated through using gatekeepers' services and applications. Vehicle manufacturers can offer new contractual terms to their own users (e.g., more competitive prices, tailored mobility services or data sharing mechanisms) in exchange for access to those applications' data. At the same time, they can develop new contractual terms and conditions with gatekeepers to provide them with the knowledge they gain from processing data the gatekeepers are prohibited from processing.

# 8. The EU AI Act

Although the EU AI Act (hereafter AI Act) applies to the use of AI in some back-end processes of vehicle manufacturing (e.g., to produce software code or to draw CAD drawings), it does not apply to the auditing of the AI systems used as components in vehicles (e.g. automatic driving assistant systems). However, vehicle manufacturers are increasingly paying attention to the development and implementation of this regulation since it forms the basis for future sector-specific AI regulations that will directly affect vehicle AI systems.

## 8.1. The EU AI Act and Type Approval Regulation

The AI Act amends multiple different existing legislative instruments in that it typically requires that previous legislation to refer to the AI Act for relevant issues. In fact, Title XII Final Provisions of the Act, amends around 11 different prior legislations. The list of these amendments can easily be found in Articles 75-82 but a few examples include amending regulations on civil aviation security,[34] regulations for the certification of agricultural and forestry vehicles,[35] regulations on approval and market surveillance of two- or three-wheel vehicles and quadricycles,[36] regulations on approval and market surveillance of motor vehicles and their trailers, and of systems, components and separate technical units intended for such vehicles.[37] The common characteristics of all these amendments is that they require prior regulations to reference the AI Act and this enables specific legislation instruments to extend and clarify for specific sectors and industries in the future.

This general approach positions the AI Act as the fundamental framework - the reference - that can be changed and adapted by different industries and sectors to match their specific context. Because the AI Act provides over-generalized amendments to prior regulations, it is appropriate to investigate the sector-specific responses and guidelines to the way the amendments are likely to be interpreted by expert groups in each sector/industry. Below, we present the example of the amendment to Regulation (EU) 2018/858 on approval and market surveillance of motor vehicles and their related systems and parts since this is one of the most relevant regulations to vehicle manufacturing (as the industry at the focus of this study). We analyze the response of the industry's expert association to exemplify how various sectors might respond to the general requirements of the AI Act and what future regulatory directions each industry can expect in relation to the AI Act.

Although it might sound counterintuitive, the automotive (as well as aviation) industry is exempt from the requirements of the AI Act in auditing its vehicle systems.[38] The reason for this exemption is multifold based on the industry's expert association (European Automobile Manufacturers Association, ACEA) to the application of the AI Act to the automotive industry and amendments to Regulation (EU) 2018/858 on approval and market surveillance of motor vehicles. Following are the major arguments presented by the European Automobile Manufacturers Association:

- First, being a highly security and safety-sensitive product, vehicles have always been subject to the utmost strict regulations on a global level. Many of the potential AI legal requirements advanced in the AI Act have already been adopted effectively into established regulations governing the automotive sector with regards to vehicle AI systems. These regulations include, for instance, ex-ante conformity assessments (e.g., Type Approval Regulation 2018/858, and General Safety Regulation) or ex-post surveillance mechanisms (e.g., conformity of production, in-service compliance, market surveillance and recalls). It is the position of the European Automobile Manufacturers Association (ACEA, 2020) that many of these already existing regulations have investigated substantially in developing practical guidelines that are under continuous progress and updates. To avoid any overlap and imposing redundant costs on manufacturers, the EU policy makers must first support the development of such practical guidelines for the existing comprehensive regulations and only then review them for possible legislative gaps.

---

[34] (Regulation (EC) No 300/2008) amended in Article 75 of the AI Act

[35] (Regulation (EU) No 167/2013) amended in Article 76 of the AI Act

[36] (Regulation (EU) No 168/2013) amended in Article 77 of the AI Act

[37] (Regulation (EU) 2018/858) amended in Article 80 of the AI Act

[38] The EU AI Act still applies to these industries if they use AI in the organizational processes. However, the industry-specific AI systems related to vehicles are already subject to strict audit requirements on a more universal level.

- Second, ACEA (2020) emphasizes that the definition of AI advanced by the AI Act and the High-Level Expert Group on Artificial Intelligence (AI HLEG) are too vague and broad. This is because they adopt a "phenomenological" description (ACEA 2020, p. 20) based on the assumption that an automated vehicle can be equated to "a mere AI device" (p. 7). The purpose of this approach, of course, has been to provide an inclusive definition that could be applied horizontally across different contexts and industries. The ACEA, however, argues that such a definition runs the risk of introducing regulations for traditional software systems and narrow AI applications. Such narrow AI systems – including, for instance, lane recognition, vehicle recognition or traffic sign recognition – are based on machine-learning algorithms trained offline and tested rigorously before being deployed as fixed software in vehicles, similar to any other software component in vehicles. Hence, to avoid misclassifying AI use cases, an accurate and narrow definition of high-risk AI systems is needed. This narrow definition should be valid only for the individual AI application and not all the AI applications across the sector, in general.

- Third, 'general AI' capable of common-sense reasoning, self-awareness, and the ability of the machine to define its own purpose is not technically feasible currently and in near future. Even learning rational systems in certificated AI systems are not involved at the current state of development for safety-critical functions. The autonomy level of today's most advanced vehicles does not even reach level 3 defined by the Society of Automobile Engineers (SAE). The current ADAS systems' level of autonomy usually corresponds to ASE's level 2. ADAS AI modules are incorporated in software that are already developed according to state-of-the-art standards and practices for functional safety and software development. Thus, they are rigorously analyzed and tested prior to deployment (ACEA, 2020). It is the ACEA's position that even if any further sector-specific regulation becomes necessary with the continuous advancements of AI technology, those regulation should "be reserved exclusively for high-risk, safety-critical AI applications related to automated driving – level 3 and above as defined by the Society of Automotive Engineers (SAE) – as long as they are incorporated into the type-approval system" (p. 2).

## 8.2. Summary of the EU AI Act

The EU claims sovereignty when it comes to privacy laws. For regulating artificial intelligence, the picture is completely different. Being one of the world's biggest social media markets, the EU's privacy laws are bound to have a consequential impact on big tech companies if they are to keep the significant European social media market. However, regulating the development and use of AI is less likely to carry a similar weight as long as the EU's contribution to research and development on AI models is remarkably lower than countries such as the U.S. It is not only the development of AI models, but also investment in such pursuits that are relatively minimal with 7 billion dollars invested by Germany (for example) between 2013 and 2022, in comparison to the U.S. 250-billion-dollar investment in AI.[39]

To reinstate its power then, the EU has adopted a particular strategic approach to AI excellence and trust. The European AI strategy aims at turning the union into the world class hub for human-centric and trustworthy AI. The approach includes three overarching initiatives: *Coordinated Plan on AI* (2021), *AI Innovation Package* (2023), and setting the landmark comprehensive *Regulatory Framework on Artificial Intelligence* (2024). Together, these initiatives are meant to build strategic leadership in high-impact sectors, strengthen investment and innovation in AI across the EU, and guarantee the safety and fundamental rights of people and businesses.

### 8.2.1. Coordinated Plan on AI

The initiative that led to the AI Act was first published in 2018 and its aim was to *coordinate* the EU Member States (as well as Norway and Switzerland). The goals of the joint Commission were to accelerate investments in AI technologies, fully and promptly implement AI strategies and programs to ensure that the EU maximizes the advantages of being an early adopter and align AI policy to remove legislative fragmentation among Member States when regulating AI.

---

[39] See the Economist, Geopolitical Monster v Brussel's Effect, Why the EU will not remain the world's digital über-regulator, 2023.

**8.2.2. AI Innovation Package**

After coordination of Member States, the first step was to encourage investment in AI research and development. The AI Innovation Package is a continuation of the EU Commission's *Large AI Grand Challenge* (a prize giving AI startups financial support and supercomputing access). The package introduces a broad range of measures to actualize the aims of the Coordinated Plan on AI initiative in concrete ways. It, for instance, amends the EuroHPC Regulation to set up AI Factories through and provide AI-dedicated supercomputers that enable fast machine learning as well as the training of large General-Purpose AI models. The concrete measures in this package include among others:

- Financial support from the Commission through Horizon Europe and the Digital Europe programs dedicated to generative AI,

- Initiatives to strengthen EU's generative AI talent pool through education, training, skilling, and reskilling activities,

- Acceleration of the development and deployment of Common European Data Spaces for access to high quality data,

- The *GenAI4EU* initiative, which aims to support the development of novel use cases and emerging applications in Europe's industrial ecosystems, as well as the public sector,

- The EIC accelerator Program and *InvestEU* that encourage public and private investments in AI start-ups and scale-ups, including through venture capital or equity support, and

- The *Alliance for Language Technologies (ALT-EDIC)* that aims to develop a common European infrastructure in language technologies to address the shortage of European languages data for the training of AI solutions, and

- The 'CitiVERSE' EDIC will apply state-of-the-art AI-tools to develop and enhance Local Digital Twins for Smart Communities, helping cities simulate and optimize processes, from traffic management to waste management.

**8.2.3. The EU Artificial Intelligence Act**

Just like the GDPR and the Digital Services Act, the AI Act follows a risk-based approach to regulating AI systems. The logic is that the accountability and management of the AI systems should match the level of risk they impose on safety, security, and fundamental rights of individuals. The AI Act recognizes four levels of risk that an AI system can pose: Level 1 (minimal risk), Level 2 (limited risk), Level 3 (high risk), Level 4 (unacceptable risk). Examples of AI systems with Level 1 risk probability include video games or filters. Examples of AI Systems with Level 2 risk probability include Deepfake or synthetic content-generating systems and chatbots. Whereas Levels 1 and 2 are subject to free use without any significant compliant requirements due to minimal or limited risk levels, AI systems with a Level 4 risk probability are completely prohibited except for very few exceptions in the context of law enforcement. Examples of AI systems with Level 4 risk probability include real-time remote biometric identification/categorization systems, or social scoring systems. These systems are prohibited because they not only pose immediate risks to individual's fundamental rights but because they promote practices that are anti-human such as untargeted surveillance and classification of individuals based on physical, psychological and behavioral metrics.

The majority of the AI systems, however, fall under the Level 3 (High) risk category and are subject to heaviest regulatory requirements. In this article, we solely focus on high-risk AI systems since the bulk of the AI Act addresses the actors in this category and because vehicle manufacturing specifically provides and deploys AI systems that fall under this category. The high-risk systems are listed through Annex I, II, and III of the AI Act:

- **Annex I:** includes all the technical methods that can introduce or perpetuate biases or safety/security issues in various AI systems. These technical methods include, machine learning approaches, logic programming and search and optimization methods. "Providers" placing on the market (or putting into service) in the EU AI systems, regardless of where they are based, and "users" of such systems within the EU are subject to heavy regulatory requirements.

- **Annex II:** is a list of all the European regulations that already exist on product safety. The legislations listed in this annex are all amendments to their predecessor directives on product safety and market surveillance that had been adopted by various Member States prior to the EU Harmonization Legislation in 2019.[40] Since each member state had established national measures to confirm the safety of various products, legal fragmentation around various manufactured products had created obstacles for free trade of products among the European states. EU's Harmonized Legislation on product safety in 2019 establishes a single body of rules for the seventy European harmonized product sectors and prohibits the Member States from adopting individual legislative measures that restrict the free movement of products. The products in this harmonized legislation include manufactured products (and not products such as food or medicine). According to Article 6 of AI Act, all the AI systems that are a product themselves or are a safety component to a product that is covered by the list of harmonized legislations in Annex II are counted as high-risk AI systems, *if* they are subject to third party safety assessment under the list of the regulations in Annex II.

    Manufactured products subject to harmonized legislations listed in Annex II include, for instance, motor vehicles and their trailers, and of systems, components and separate technical units intended for such vehicles, machinery, toys, personal watercraft, to lifts and safety components for lifts, radio equipment, two- or three-wheel vehicles, agricultural and forestry vehicles, motor vehicles and their trailers, and systems, components and separate technical units, diagnostic medical devices, etc.

- **Annex III**: whereas Annex I addresses technical *methods*, and Annex II addresses high-risk *products* and *components*, Annex III focuses on high-risk *contexts*. Annex III emphasizes that although an AI system might not be high-risk per se (e.g., it does not use complex machine learning approaches or logic programming), it still can pose risks to individuals health, security or fundamental rights because those contexts tend to facilitate or block individuals' access to vital life resources (e.g., education, employment), or affect a large group of individuals (e.g., critical infrastructures). These contexts include: any context where the use of biometrics systems is allowed, critical infrastructure, education and vocational training, employment, access to essential private services and essential public services and benefits (e.g., systems used to confirm individuals loan creditworthiness, or systems intended to evaluate and classify emergency calls), law enforcement, migration, justice system.

| EU Artificial Intelligence Act (AIA) ||
|---|---|
| **Problem** | **Instrument** |
| **Legal** ||
| • Extremely **scattered attempts** that aimed at regulating the **wide-span use** of AI systems in various aspects of life | • Regulate AI **development**, **use**, and **distribution** on the **Union level** |

---

[40] Regulation (EU) 2019/1020 on market surveillance and compliance of products

| | |
|---|---|
| ● **Lack** of a **unified supervisory** authority | ● **Create** an **EU-level registry database** for registering high-risk AI systems by providers (to be **supervised** by the **EU Commission**) |
| ● Nascency of the AI applications and possible negative effects of regulations on innovation capabilities | ● Develop a **tiered approach** to regulations<br>● Encourage development of AI technologies by providing legal certainty and ensure that no obstacle to the cross-border provision of AI-related services and products emerge |
| ● **Lack** of **clarity** on the accountability of **different actors** across the AI technologies value chain | ● Set **proportionate obligations** on **all value chain participants** (providers, importers, distributors, users) |
| ● The **potential** of AI systems to **identify**, **classify**, and **discriminate** in particularly **sensitive aspects** of **life** such as safety, social benefits, or employment. | ● Designing the **assessment** and **mitigation requirements** based on EU's Charter of **Fundamental Right** including:<br>– Right to human dignity (Article 1),<br>– Respect for private life and protection of personal data (Articles 7 and 8),<br>– Right to nondiscrimination (Article 21),<br>– Right to equality between women and men (Article 23),<br>– Right to freedom of expression (Article 11),<br>– Right to freedom of assembly (Article 12),<br>– Right to effective remedy and to a fair trial,<br>– Rights of defense and the presumption of innocence (Articles 47 and 48),<br>– The rights of a number of special groups, such as the workers' rights to fair and just working conditions (Article 31),<br>– Right to a high level of consumer protection (Article 28),<br>– The rights of the child (Article 24),<br>– The right to integration of persons with disabilities (Article 26),<br>– The right to a high level of environmental (Article 37) |
| **Technical** | |
| ● **Lack** of public **trust** in the use of AI due to **lack** of **control and enforcement mechanisms** | ● Introducing a European **coordination mechanism**, and facilitating **audits** of the AI systems with **new requirements** for **documentation**, **traceability** and **transparency** |
| ● The problem of AI systems **black box** | ● **Mandate information provision** in the form of **documentation** and instructions of use with concise and clear information, including in relation to possible risks to fundamental rights and discrimination<br>● **Mandate information provision** on the general **characteristics**, **capabilities** and **limitations** of the system, algorithms, data, training, testing and validation processes used as well as documentation on the relevant risk management system |
| ● The **moving frontier** in AI technologies | ● **Mandate** the conduct of a **new conformity assessment** whenever a **change occurs** which may affect the compliance of the system with this Regulation or when the intended purpose of the system changes |
| **Economic** | |
| ● Considerably **high costs** of **compliance** requirement | ● Adopt a **tiered approach** toward compliance |
| ● Negative effects on **SMEs** | ● Create of **regulatory sandboxes**<br>● Create obligation to consider SMEs interests when setting fees |
| ● **Duplication** of already **existing** regulations (e.g., the New Legislative Framework approach, Machinery Regulation) | ● Consider compliance assessment with AI Act as **part of conformity assessment** with the **existing regulations** |

*Table 7. Summary of the EU AI Act (AIA)*

As mentioned previously, many AI vehicle systems are already heavily regulated, and the AI Act does not apply to vehicle systems. So, the question remains as to why vehicle manufacturers need to keep an eye on the AI Act. The answer is in the ACEA's critical remark about the definition of AI in the context of automated vehicles. The ACEA criticizes the AI Act's definition of AI and argues that the Act only recounts certain static characteristics to be descriptive of AI systems and offers an all-encompassing description (intended to be applicable horizontally to all AI systems). The Act's over generalized definition emphasizes AI as a single, discernable thing, or a discrete phenomenon or a set of technologies. However, the ACEA recognizes that if we understand automated vehicles not as an automated device but as an ever-evolving frontier of computational and processing capabilities and scope of application (Berente et al. 2021), then we recognize the need for further regulations in a near future when vehicles approach SAE's level 3 of automation. To this point ACEA identifies a few areas where both the automotive society and policy makers will have to continuously monitor.

### 8.3. Interaction with the GDPR: Opportunity

- **Records of processing data (ROPA) and data provenance:** vehicle manufacturers need to audit functional and operational safety not only during the design and development processes of the automated driving system but also during the vehicle life cycle (ACEA 2020). Performing such audits requires records that provide clear information about the storage, documentation, and retention of the training data sets and methods. Vehicle manufacturers have a vantage point when it comes to having a clearer record of their data storage and processing activities. This is because, as organizations that control and process large amounts of data, they have been subject to mandatory GDPR requirements ((Article 25(2)) to keep clear records of their processing activities (ROPAs). They have, thus, created the fundamental mechanisms and capabilities for data retention and provenance required to audit AI systems. One opportunity is, therefore, that vehicle manufacturers can more easily expand on these fundamental mechanisms and capabilities compared with organizations that have little experience with developing data retention processes and capabilities. As the project manager for implementation of Data Act and AI Act in one of the vehicle manufacturers in our study elaborated:

    *"Now we are going to register every AI system we have, even though the AI Act does not apply to us at the moment. We got input from another OEM who's ahead of us in this, they started way before the AI Act was passed. So, now we, too, will register all the AI systems that we have. It is part of master data management and knowing where you have all the data and tagging it. You now need to use where you are using AI and how".*

    Not all organizations have been subject to mandatory record keeping. However, those that have voluntarily engaged in keeping such processes can also benefit from a similar vantage point. Data documentation and retention is significant for arranging and validating the correct cycles around the model training and building the vital capabilities to manage AI models life cycle (Weber et al. 2022). Such capabilities, in turn, help organizations to provide a proper data foundation, understand and cope with data dependency and continuously adjust AI systems as the data evolves (Weber et al. 2022). These capabilities contribute to a more explainable and scrutable (Hassan and Ojala 2024; Asatiani et al. 2024), and therefore ethical AI development (Weber et al. 2022; Berente et al. 2021; Pushkarna et al 2023).

- **Impact assessments requirement and preparedness: data controller under the GDPR vs. user under the EU AI Act:** Vehicle manufacturers collect data categories and process them in ways that pose high risks to the fundamental rights of natural persons defined in Article 35 of the GDPR. These high-risk processes include, for instance, large-scale profiling of users to predict future trendsetting car buyers or data processed by smart meters or IoT applications, or the use of new technologies, or the novel application of the old ones such as artificial intelligence and machine learning in connected and autonomous vehicles. Vehicle manufacturers are therefore required to conduct thorough data processing impact assessments (DPIAs). Additionally, vehicle manufacturing represents a context where most advanced data processing and automated technologies are incorporated in a safety and security sensitive product such as a vehicle. Many of these advanced technologies are subject to constant universal assessments and auditing requirements (e.g., General Safety Regulation, or Type Approval Regulation) which has enabled them to navigate a smooth path in compliance with the GDPR. Similarly, conducting constant DPIAs for the sake of staying compliant with the GDPR enables vehicle manufacturers to be more prepared for compliance with other regulatory instruments such as the EU AI Act (or equivalent industry-specific regulations on AI systems) which focus heavily on issues of data handling.

One example is that the GDPR considers vehicle manufacturers as a data controller under the GDPR and subject to strictest data related assessments. However, according to the EU AI Act a vehicle manufacturer is considered as an AI user and therefore subject to more limited mandatory assessment requirements. Hence, it is likely that the assessment efforts for compliance with the GDPR potentially work as fundamentally useful steps for compliance with AI- related regulations.

## 8.4. Interaction with the GDPR: Tension

- **Informed safety: Information provision obligations for IoT data vs. personal data:** another issue is the information provision requirements in relation to the design of the vehicles' operational domain. To understand AI in terms of specific use cases rather than a general phenomenon applicable everywhere (Weber et al. 2022), or a mere device (ACEA 2020; Berente et al. 2021) means to evaluate its ethical aspects in relation to the context of application (see Martin 2016; Martin 2023). The ACEA (2020) emphasizes that, rather than the technology used, regulating AI systems should be based on understanding the degree of automation in relation to vehicle behavior. The concept of Operational Design Domain in vehicle manufacturing is based on this contextual understanding. Simply put, Operational Design Domain[41] is defined as the vehicle's capabilities and limitations (e.g., speed limit, switch to manual mode) in a given context (e.g., on a slippery road, when the driver's heart rate is increased). Operational Design Domain refers to a vehicle's operation domain in a given context. Operational Design Domain is directly related to an ethical design of automated vehicles since such a design requires that the user understands the 'conditions' in which an automated vehicle is capable of operating safely, a notion known as "informed Safety" (Khastgir et al. 2018).

    Informed safety is addressed at UN level by WP29 – that relates to informing the customers about the capabilities and limitations of an automated driving system in a given context – and WP1 – that relates to new driver training requirements in the future. Informed safety is thus aiming at emphasizing users' actual agency through not only opening the black box of automated systems and informing them about the design limits and capabilities of systems, but also through informing users on the dynamism of the environments in which automated technologies operate (Zhang et al. 2021).

    In the context of automated vehicles, two examples can be discussed in relation to informed safety. When, for instance, the heart rate or the eye movements of the driver suggest a lowered focus on driving, the vehicle suggests the driver switches to automated driving. To provide the user with such autonomy, however, sensitive data such as users' health information needs to be collected. In this example, the collected data is not for the safe functioning of the vehicle, though. The vehicle is capable of driving safely without this data. Thus, the purpose of data collection is not completely straightforward, and as explained below, this will have implications for compliance with the GDPR (Mulder and Vellinga 2018).

    Another example is when it is necessary to communicate with other road users that a driver of a vehicle is facing possible safety hazards and can pose a threat to safety of the vehicles. This communication requires sharing (possibly sensitive) information about the vehicle's status and intention and the drivers' health condition. In both these examples, Article 13 and 14 of the GDPR require the data holder to provide the user with certain information about, for instance, the purpose of data collection, identity of data holder, period of data storage, disclosure to other recipients, rights to erasure of data, etc.

    Informed safety, thus, emphasizes an ethical AI development by aiming at giving the users the autonomy of decision making to both make decisions or choose to not to make decisions (see Floridi et al. 2018). However, the ACEA also emphasizes that such information provision about

---

[41] Legislation developed/adopted within the framework of the Working Party on Automated/Autonomous and Connected Vehicles – GRVA (WP29/WP1).

the continuous and large flows of connected vehicle's data might be impossible or require disproportionate effort. In these conditions, future regulations should both consider GDPR's exceptions to information provision obligations (Article 14 (5) b) and establish certain standards or guidelines as to how the information provision obligations would look like within AI systems.

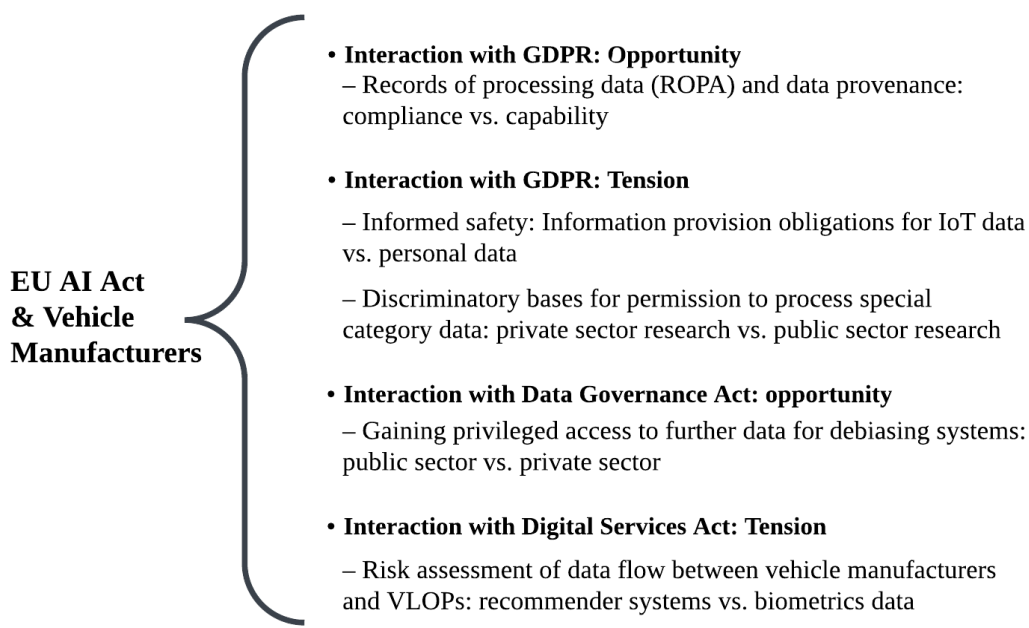

**EU AI Act & Vehicle Manufacturers**

- **Interaction with GDPR: Opportunity**
  – Records of processing data (ROPA) and data provenance: compliance vs. capability

- **Interaction with GDPR: Tension**
  – Informed safety: Information provision obligations for IoT data vs. personal data
  – Discriminatory bases for permission to process special category data: private sector research vs. public sector research

- **Interaction with Data Governance Act: opportunity**
  – Gaining privileged access to further data for debiasing systems: public sector vs. private sector

- **Interaction with Digital Services Act: Tension**
  – Risk assessment of data flow between vehicle manufacturers and VLOPs: recommender systems vs. biometrics data

*Figure 6. The Interaction of EU Data AI Act with Other Regulations*

- **Discriminatory bases for permission to process special category data: private sector research vs. public sector research:** the ACEA underscores the importance of data sharing for addressing the problem of bias in AI systems. Improving AI systems for certain legitimate goals may require processing data that are currently prohibited under Article 9 of the GDPR as special data categories. One of these categories is for instance biometrics data. The processing of these special categories of data is subject to exceptions under specific circumstances such as research in public interest (GDPR, Article 9(2-j); GDPR, Article 89(2)). Although conducting research on driver and road safety in automated driving could also be interpreted as research for public interest, and despite Recital 159 GDPR stipulating that privately funded research also has recourse to these privileges, private sector research is usually denied such privileges. Similarly, data sharing among countries where a vehicle manufacturer has operations should be facilitated and regulated, according to the ACEA. This tension associated with limited data processing and more biased systems is a relatively more important issue that future legislation will likely address.

### 8.5. Interaction with the EU Data Governance Act: Opportunity

- **Gaining privileged access to further data for debiasing systems: public sector vs. private sector:** The ACEA's position paper was published before the passing of many EU's data sharing regulations discussed here. The EU Data Governance Act and Data Act specifically regulate international data sharing with the aim of expediting such pursuits. Compliance requirements with current and future AI regulations in the automotive industry could serve as a basis for benefiting from the provisions in, for instance, Data Governance Act. For instance, debiasing systems require access to representative and appropriate-size data that are often counted as sensitive and special category data (e.g., biometrics, health records). Public sector bodies seem to be privileged in accessing special category data and are holders of large amounts of sensitive/protected data. The Data Governance Act regulates the sharing of data held by public sector bodies and introduces the concept of trusted and regulated intermediaries to provide novel and trusted research environments. This should make it interesting for vehicle manufacturers to seek opportunities for using such environments and legal bases to creatively develop new cross-sectoral research collaborations.

## 8.6. Interaction with the Digital Services Act: Tension

- **Risk Assessments of data flow between vehicle manufacturers and online platforms:** Perhaps one of the most important interactions of the AI Act (or other similar regulations related to AI systems for vehicle manufacturers) would be with the Digital Services Act. The AI Act adopts a risk-based approach to auditing intelligent systems. Vehicle manufacturing entails some of the most high-risk automated areas related to critical infrastructures that could put the life and health of citizens at risk (e.g., transport) or threaten their fundamental rights (e.g., remote biometric identification systems). These high-risk areas will be subject to the strictest auditing requirements. At the same time, VLOPs employ algorithmic recommender systems for targeted advertising. The Digital Services Act points to the significant effects of these algorithmic recommender systems on identifying and targeting the users and the risks they could impose on fundamental rights (Recital 55 & 70). Since vehicles collect special categories of data such as biometrics, they might be required to be transparent in terms of the data exchange cases with VLOPs. Thus, although the Digital Services Act does not directly apply to vehicle manufacturers, vehicle manufacturers' relationship with VLOPs (with highest liability potential in these sensitive areas) could potentially affect the liability risks and auditing of vehicle systems under the AI Act.

# 9. Data and AI Regulations as a Wave Set

Clearly different data and AI regulations are not independent of each other. Rather, organizations can expect these waves of legislation - rooted in the past and building on and interacting with each other - to be the new norm. As data and AI related technologies evolve at an ever-more-rapid pace, the regulations of these technologies will attempt to keep pace.

These wave sets of data and AI legislation have implications for the management of organizations. First, it is increasingly difficult for organizations to take a discrete, compliance-oriented view of each legislation. Compliance with any given legislation now often requires extensive transformation of technological infrastructures and governance frameworks and practices. Further, sometimes these legislative instruments interact and even contradict each other. For instance, the tension points between different regulations (GDPR and the Data Act) place unprecedented requirements on vehicle manufacturers to rethink their identification systems and contracts, and trace legal bases for processing/sharing different categories of data. Organizations can no longer afford to think about data and AI legislation in isolation, but need to understand how these legislations act as a wave set - ongoing and never ending.

However, accompanying these challenges, the data and AI legislation wave set also presents multiple opportunities to generate value. As a result of developing technological and governance infrastructure, organizations may find advantageous connections with other legislative instruments, and come up with creative ways to gain synergies across the legislations and mindfully identify new ways where the organizations may generate value. An example is when the interaction between the Digital Markets Act and Data Act allows manufacturers to form innovative contractual agreements for data processing/sharing, or when the interaction between Data Governance Act and Data Act allow manufacturers to form more trustworthy research consortia and benefit from the interoperability capabilities that are created as a result of data exchange. Figure 7 illustrates some of these opportunities across the data and AI legislation wave set.

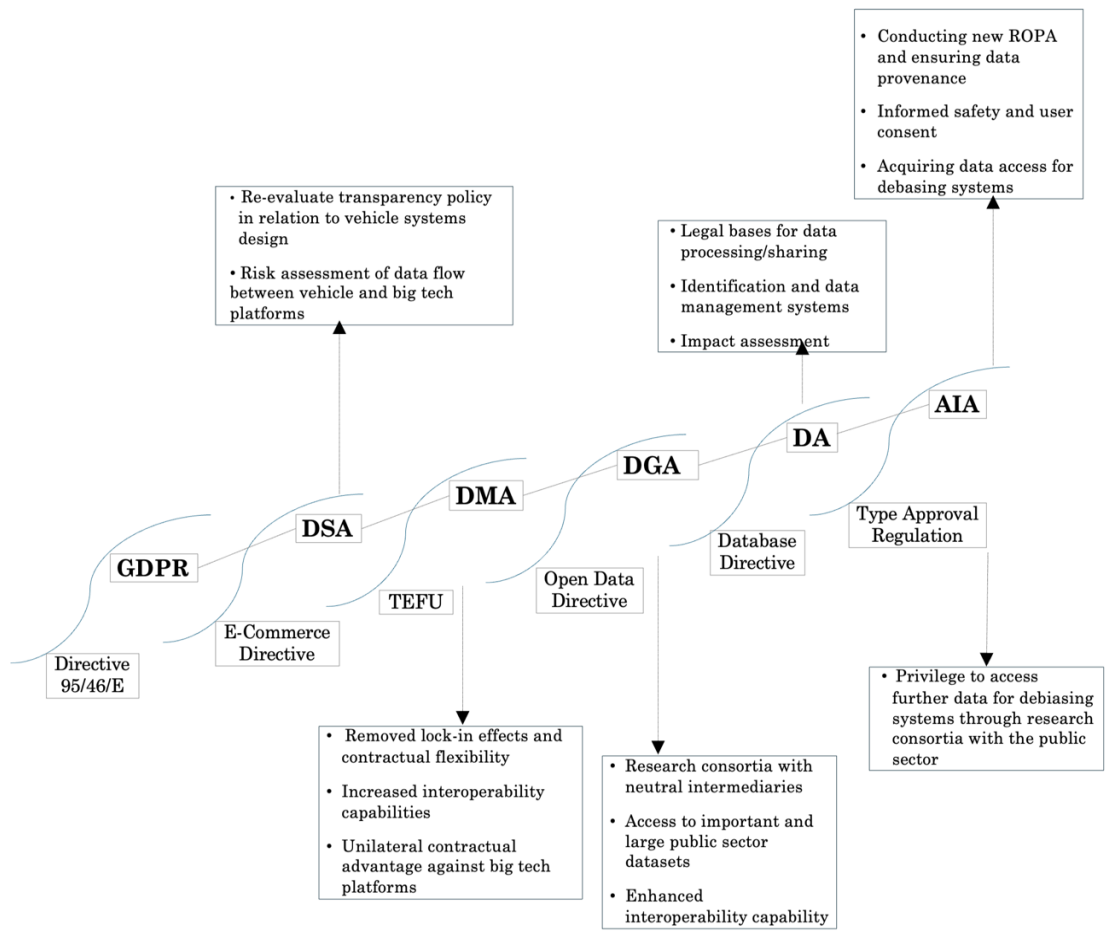

*Figure 7. Wave Set of EU Data and AI Regulations*

Figure 7 also exemplifies an important point about the historical progression of regulations in the EU, i.e., each regulation is preceded by other legislative instruments. Paying close attention to the amended points and the *reason* these points have been modified and adapted can help organizations predict the future direction of data and AI-related regulatory instruments. For instance, the Digital Services Act modifies liability conditions in the E-Commerce Directive and emphasizes that appropriation — i.e., presenting content in a way that they seem as part of a product/service/platform — will imply an active role in deception or illegal activity. One of the lessons learned for vehicle manufacturers could then be paying attention to the way they formulate privacy policies with big tech service providers whose applications and core platform services are presented as part of a product such as vehicles and thereby understood by users as inseparable from them.

In the end, a key takeaway from this analysis is that managers can no longer afford to relegate thinking about this legislation to compliance professionals only. Certainly, legal and compliance professionals need to be involved, but data and AI is now a strategic imperative for every type of organization - even canonical industrial organizations like vehicle manufacturers. Managers, engineers, and any professionals with a strategic view in organizations need to understand how these legislations constrain and enable the generation of value going forward. In the digital age, not only do effective managers need to understand wave upon wave of technologies themselves, but also the accompanying wave set of regulations.

# Table of Legislation

**International Conventions, and Regulations**

- United Nations, Regulation No. 157. Uniform provisions concerning the approval of vehicles with regard to Automated Lane Keeping Systems Geneva, 24 June 2020.

- United Nations, General Assembly, International Covenant on Civil and Political Rights, 16 December 1966, United Nations, Treaty Series, vol. 999, p. 171.

- UN Legislation developed/adopted within the framework of the Working Party on Automated/ Autonomous and Connected Vehicles – GRVA (WP29/WP1).

**United State of America Regulations**

- U.S. Congress. (1958) United States Code: Copyright Office, 17 U.S.C. §§ 201-216.

- 47 U.S.C. 230; Communications Decency Act, 1996.

**European Legislation**

**- EU Primary Legislation**

- European Union, Charter of Fundamental Rights of the European Union, 26 October 2012, 2012/C 326/02.

- Council Regulation (EC) No 1/2003 of 16 December 2002 on the implementation of the rules on competition laid down in Articles 81 and 82 of the Treaty.

- Consolidated version of the Treaty on the Functioning of the European Union, OJ C 326, 26.10.2012.

**- EU Secondary Legislation**

- Regulation (EU) 2016/679 of the European Parliament and of the Council of 27 April 2016 on the protection of natural persons with regard to the processing of personal data and on the free movement of such data, and repealing Directive 95/46/EC (General Data Protection Regulation) OJ L 119, 4.5.2016.

- Regulation (EU) 2022/868 of the European Parliament and of the Council of 30 May 2022 on European data governance and amending Regulation (EU) 2018/1724 (Data Governance Act) PE/85/2021/REV/1 OJ L 152, 3.6.2022.

- Regulation (EU) 2022/2065 of the European Parliament and of the Council of 19 October 2022 on a Single Market for Digital Services and amending Directive 2000/31/EC (Digital Services Act) PE/30/2022/REV/1 OJ L 277, 27.10.2022.

- Regulation (EU) 2022/1925 of the European Parliament and of the Council of 14 September 2022 on contestable and fair markets in the digital sector and amending Directives (EU) 2019/1937 and (EU) 2020/1828 (Digital Markets Act) PE/17/2022/REV/1 OJ L 265, 12.10.2022.

- Regulation (EU) 2023/2854 of the European Parliament and of the Council of 13 December 2023 on harmonized rules on fair access to and use of data and amending Regulation (EU) 2017/2394 and Directive (EU) 2020/1828 (Data Act) PE/49/2023/REV/1 OJ L, 2023/2854, 22.12.2023.

- Regulation (EU) 2024/1689 of the European Parliament and of the Council of 13 June 2024 laying down harmonized rules on artificial intelligence and amending Regulations (EC) No 300/2008, (EU) No 167/2013, (EU) No 168/2013, (EU) 2018/858, (EU) 2018/1139 and (EU) 2019/2144 and Directives 2014/90/EU, (EU) 2016/797 and (EU) 2020/1828 (Artificial Intelligence Act), PE/24/2024/REV/1 OJ L, 2024/1689, 12.7.2024.

- Regulation (EU) 2018/1725 of the European Parliament and of the Council of 23 October 2018 on the protection of natural persons with regard to the processing of personal data by the Union institutions, bodies, offices and agencies and on the free movement of such data, and repealing Regulation (EC) No 45/2001 and Decision No 1247/2002/EC.

- Regulation (EU) 2018/858 of the European Parliament and of the Council of 30 May 2018 on the approval and market surveillance of motor vehicles and their trailers, and of systems, components and separate technical units intended for such vehicles, amending Regulations (EC) No 715/2007 and (EC) No 595/2009 and repealing Directive 2007/46/EC.

- Regulation (EC) No 300/2008 of the European Parliament And Of The Council of 11 March 2008 on common rules in the field of civil aviation security and repealing Regulation (EC) No 2320/2002, OJ L 97, 9.4.2008

- Regulation (EU) No 167/2013 of the European Parliament and of the Council of 5 February 2013 on the approval and market surveillance of agricultural and forestry vehicles OJ L 60, 2.3.2013.

- Regulation (EU) No 168/2013 of the European Parliament and of the Council of 15 January 2013 on the approval and market surveillance of two- or three-wheel vehicles and quadricycles OJ L 60, 2.3.2013.

- Regulation (EU) 2018/858 of the European Parliament and of the Council of 30 May 2018 on the approval and market surveillance of motor vehicles and their trailers, and of systems, components and separate technical units intended for such vehicles, amending Regulations (EC) No 715/2007 and (EC) No 595/2009 and repealing Directive 2007/46/EC, PE/73/2017/REV/1 OJ L 151, 14.6.2018.

- Regulation (EU) 2019/1020 of the European Parliament and of the Council of 20 June 2019 on market surveillance and compliance of products and amending Directive 2004/42/EC and Regulations (EC) No 765/2008 and (EU) No 305/2011, PE/45/2019/REV/1 OJ L 169, 25.6.2019

**- EU Directives**

- Directive (EU) 2016/680 of the European Parliament and of the Council of 27 April 2016 on the protection of natural persons with regard to the processing of personal data by competent authorities for the purposes of the prevention, investigation, detection or prosecution of criminal offenses or the execution of criminal penalties, and on the free movement of such data, and repealing Council Framework Decision 2008/977/JHA.

- Directive 95/46/EC of the European Parliament and of the Council of 24 October 1995 on the protection of individuals with regard to the processing of personal data and on the free movement of such data.

- Directive 2009/22/EC of the European Parliament and of the Council of 23 April 2009 on injunctions for the protection of consumers' interests OJ L 110, 1.5.2009.

- Directive 96/9/EC of the European Parliament and of the Council of 11 March 1996 on the legal protection of databases OJ L 77, 27.3.1996.

- Directive 2003/98/EC of the European Parliament and of the Council of 17 November 2003 on the re-use of public sector information OJ L 345, 31/12/2003.

- Directive 2013/37/EU of the European Parliament and of the Council of 26 June 2013 amending Directive 2003/98/EC on the re-use of public sector information Text with EEA relevance OJ L 175, 27.6.2013.

- Directive 2000/31/EC of the European Parliament and of the Council of 8 June 2000 on certain legal aspects of information society services, in particular electronic commerce, in the Internal Market ('Directive on electronic commerce') OJ L 178, 17.7.2000.

- Directive (EU) 2019/1024 of the European Parliament and of the Council of 20 June 2019 on open data and the re-use of public sector information, PE/28/2019/REV/1, OJ L 172, 26.6.2019.

**- EU Official Documents**

- Article 29 Data Protection Working Party, Guidelines on Personal Data Breach Notification under Regulation 2016/679, Adopted on 3 October 2017 as last Revised and Adopted on 6 February 2018.

- Commission Staff Working Document Evaluation Of Directive 96/9/EC on the Legal Protection of Databases, https://digital-strategy.ec.europa.eu/en/library/staff-working-document-and-executive-summary-evaluation-directive-969ec-legal-protection-databases.

- Public Consultation on Data Act and Amended Rules on the Legal Protection of Databases, https://digital-strategy.ec.europa.eu/en/library/public-consultation-data-act-summary-report#:~:text=The%20Commission%20held%20a%20public,proposal%20for%20a%20Data%20Act.

- Commission Staff Working Document Evaluation Of Directive 96/9/EC on the Legal Protection of Databases, https://digital-strategy.ec.europa.eu/en/library/staff-working-document-and-executive-summary-evaluation-directive-969ec-legal-protection-databases.

- Commission Staff Working Document Evaluation of Directive 96/9/EC on the legal protection of databases, Brussels, 25.4.2018 SWD(2018) 146 final, https://edz.bib.uni-mannheim.de/edz/pdf/swd/2018/swd-2018-0146-en.pdf.

- European Commission, DG Internal Market, First evaluation of the Directive 96/9/EC on the legal protection of databases (2005).

- Commission Staff Working Document Impact Assessment Report Accompanying the Document Proposal For A Regulation of the European Parliament and of The Council on European Data Governance (Data Governance Act), 2020, Brussels, 25.11.2020, SWD(2020) 295 final, https://eur-lex.europa.eu/legal-content/EN/TXT/?uri=CELEX%3A32013L0037.

- Commission Staff Working Document Impact Assessment Accompanying The Document Proposal For A Regulation Of The European Parliament And Of The Council on a Single Market For Digital

- Services (Digital Services Act) and amending Directive 2000/31/EC, Brussels, 15.12.2020 SWD(2020) 348 final,
https://digital-strategy.ec.europa.eu/en/library/impact-assessment-digital-services-act.

- European Commission's Summary report on the Public Consultation On Data Act And Amended Rules On The Legal Protection Of Databases, 2021.
https://digital-strategy.ec.europa.eu/en/library/public-consultation-data-act-summary-report#:~:text=The%20Commission%20held%20a%20public,proposal%20for%20a%20Data%20Act.

- Communication From The Commission To The European Parliament And The Council on Data protection as a pillar of citizens' empowerment and the EU's approach to the digital transition - two years of application of the General Data Protection Regulation, Brussels, 24.6.2020, SWD(2020) 115 final,
https://eur-lex.europa.eu/legal-content/EN/TXT/HTML/?uri=CELEX:52020DC0264.

- Commission Staff Working Document Impact Assessment Report Accompanying The Document Proposal For A Regulation Of The European Parliament And Of The Council on European data governance (Data Governance Act), SWD/2020/295 final,
https://eur-lex.europa.eu/legal-content/GA/TXT/?uri=CELEX:52020SC0295.

- Proposal for a Regulation by the Council and the European Parliament introducing a new competition tool, Ref. Ares(2020)2877634 - 04/06/2020.
https://eur-lex.europa.eu/legal-content/EN/ALL/?uri=PI_COM%3AAres%282020%292877634

- Publications Office of the European Union final report on Competition Policy for the Digital Era, 2019,
https://op.europa.eu/en/publication-detail/-/publication/21dc175c-7b76-11e9-9f05-01aa75ed71a1/language-en.

-European automobile Manufacturer's Association (ACEA). 2020. Position Paper on Artificial Intelligence in the Automobile Industry,
https://www.acea.auto/files/ACEA_Position_Paper-Artificial_Intelligence_in_the_automotive_industry.pdf.

# References

- Acquisti, A., 2023. The economics of privacy at a crossroads. *Economics of Privacy. University of Chicago Press.*

- Broomfield, H., 2023. Where is Open Data in the Open Data Directive?. *Information Polity*, (Preprint), pp.1-14.

- Asatiani, A., Malo, P., Nagbøl, P.R., Penttinen, E., Rinta-Kahila, T. and Salovaara, A., 2021. Sociotechnical Envelopment of Artificial Intelligence: An Approach to Organizational Deployment of Inscrutable Artificial Intelligence Systems. *Journal of the association for information systems*, *22*(2), pp.325-352.

- Berente, N., Gu, B., Recker, J. and Santhanam, R., 2021. Managing artificial intelligence. *MIS Quarterly*, *45*(3).

- Casolari, F., Buttaboni, C. and Floridi, L., 2023. The EU Data Act in context: a legal assessment. *International Journal of Law and Information Technology*, *31*(4), pp.399-412.

- Cennamo, C., Kretschmer, T., Constantinides, P., Alaimo, C. and Santaló, J., 2023. Digital Platforms Regulation: An Innovation-Centric View of the EU's Digital Markets Act. *Journal of European Competition Law & Practice*, *14*(1), pp.44-51.

- Cichy, P., Salge, T.O. and Kohli, R., 2021. Privacy Concerns and Data Sharing in the Internet of Things: Mixed Methods Evidence from Connected Cars. *MIS Quarterly*, *45*(4).

- Cohen, J.E., 2019. *Between Truth and Power*. Oxford University Press.

- Crain, M., 2018. The Limits of Transparency: Data Brokers and Commodification. *New Media & Society*, *20*(1), pp.88-104

- Crémer, J., Crawford, G.S., Dinielli, D., Fletcher, A., Heidhues, P., Schnitzer, M. and Morton, F.M.S., 2023. Fairness and Contestability in the Digital Markets Act. *Yale Journal on Regulation*. (40), pp.973.

- Economist, *Geopolitical Monster v Brussel's Effect, Why the EU Will not remain the World's Digital Über-Regulator*, 2023. Accessed June 2024, https://www.economist.com/europe/2023/09/21/why-the-eu-will-not-remain-the-worlds-digital-uber-regulator.

- Finck, M. and Pallas, F., 2020. They Who Must Not Be Identified—Distinguishing Personal from Non-Personal Data under the GDPR. *International Data Privacy Law*, *10*(1), pp.11-36.

- Floridi, L., Cowls, J., Beltrametti, M., Chatila, R., Chazerand, P., Dignum, V., Luetge, C., Madelin, R., Pagallo, U., Rossi, F. and Schafer, B., 2018. AI4People—an ethical framework for a good AI society: opportunities, risks, principles, and recommendations. *Minds and Machines*, *28*, pp.689-707.

- Gill, D. (2022). The Data Act Proposal and the Problem of Access to In-Vehicle Data and Resources. SSRN Electronic Journal. doi: https://doi.org/10.2139/ssrn.4115443.

- Hasan, R. and Ojala, A., 2024. Managing Artificial Intelligence in International Business: Toward a Research Agenda on Sustainable Production and Consumption. *Thunderbird International Business Review.*

- Hodapp, D. and Hanelt, A., 2022. Interoperability in the Era of Digital Innovation: An Information Systems Research Agenda. *Journal of Information Technology*, *37*(4), pp.407-427.

- Jacobides, M.G., Cennamo, C. and Gawer, A., 2024. Externalities and Complementarities in Platforms And Ecosystems: From Structural Solutions to Endogenous Failures. *Research Policy*, *53*(1), 104906.

- Kerber, W., 2023. Governance of IoT Data: Why the EU Data Act Will Not Fulfill Its Objectives. *GRUR International*, *72*(2), pp.120-135.

- Khastgir, S., Birrell, S., Dhadyalla, G. and Jennings, P., 2018. Calibrating trust through knowledge: Introducing the concept of informed safety for automation in vehicles. *Transportation research part C: emerging technologies*, *96*, pp.290-303.

- Kubicek, H., Cimander, R. and Scholl, H.J., 2011. *Organizational interoperability in e-government: lessons from 77 European good-practice cases* (pp. I-XIV). Heidelberg: Springer.

- Larouche, P. and de Streel, A., 2021. The European digital Markets Act: A Revolution Grounded on Traditions. *Journal of European Competition Law & Practice*, *12*(7), pp.542-560.

- Martin, K., 2016. Understanding privacy online: Development of a social contract approach to privacy. *Journal of business ethics*, *137*, pp.551-569.

- Martin, K., 2016. Understanding Privacy Online: Development of a Social Contract approach to Privacy. *Journal of Business Ethics*, *137*, pp. 555-569.

- Martin, K., 2023. Platforms, Privacy & The Honeypot Problem. *Harvard Journal of Law & Technology, Forthcoming*.

- Marty, F.M. and Warin, T., 2020. Digital Platforms' Information Concentration: From Keystone Players to Gatekeepers. papers.ssrn.com.

- McCarthy, M., Seidl, M., Mohan, S., Hopkin, J., Stevens, A. and Ognissanto, F., 2017. Access to in-Vehicle Data and Resources. *Study Commissioned by European Commission* CPR2419. Brussels, p.10. Retreived at https://www.michaellsena.com/wp-content/uploads/2014/03/2017-05-access-to-in-vehicle-data-and-resources.pdf.

- Williams, M., Nurse, J.R. and Creese, S. 2016. August. The Perfect Storm: The Privacy Paradox and the Internet-of-Things. In *2016 11th International Conference on Availability, Reliability and Security (ARES)*, IEEE, pp. 644-652.

- Mulder, T. and Vellinga, N.E., 2021. Exploring Data Protection Challenges of Automated Driving. *Computer Law & Security Review*, *40*, 105530.

- Neisse, R., Baldini, G., Steri, G., Miyake, Y., Kiyomoto, S. and Biswas, A.R. 2016. An Agent-Based Framework for Informed Consent in the Internet of Things. In *the 23rd International Conference on Telecommunications (ICT), Greece*.

- Oberländer, A.M., Röglinger, M., Rosemann, M. and Kees, A., 2018. Conceptualizing business-to-thing interactions–A Sociomaterial Perspective on the Internet of Things. *European Journal of Information Systems*, *27*(4), pp.486-502.

- Politou, E., Alepis, E. and Patsakis, C., 2018. Forgetting Personal Data and Revoking Consent under the GDPR: Challenges and Proposed Solutions. *Journal of cybersecurity*, *4*(1), pp. 1-20.

- Rantos, K., Drosatos, G., Kritsas, A., Ilioudis, C., Papanikolaou, A. and Filippidis, A.P., 2019. A Blockchain-Based Platform for Consent Management of Personal Data Processing in the IoT Ecosystem. *Security and Communication Networks*, *2019*, pp.1-15.

- Richter, H., 2023. Looking at the Data Governance Act and Beyond: How to Better Integrate Data Intermediaries in the Market Order for Data Sharing. *GRUR International*, *72*(5), pp.458-470.

- Saadatmand, F., Lindgren, R. and Schultze, U., 2019. Configurations of Platform Organizations: Implications for Complementor Engagement. *Research policy*, *48*(8), 103770.

- Srnicek, N., 2017. *Platform Capitalism*. John Wiley & Sons.

- Turillazzi, A., Taddeo, M., Floridi, L. and Casolari, F., 2023. The Digital Services Act: An Analysis of Its Ethical, Legal, and Social Implications. *Law, Innovation and Technology*, *15*(1), pp.83-106.

- Uzunca, B., Sharapov, D. and Tee, R., 2022. Governance Rigidity, Industry Evolution, and Value Capture in Platform Ecosystems. *Research policy*, *51*(7), 104560.

- Von Ditfurth, L. and Lienemann, G., 2022. The Data Governance Act: –Promoting or Restricting Data Intermediaries?. *Competition and Regulation in Network Industries*, *23*(4), pp.270-295.

- Weber, M., Engert, M., Schaffer, N., Weking, J. and Krcmar, H., 2023. Organizational capabilities for ai implementation—coping with inscrutability and data dependency in ai. *Information Systems Frontiers*, *25*(4), pp.1549-1569.

- Zhang, Z., Yoo, Y., Lyytinen, K. and Lindberg, A., 2021. The unknowability of autonomous tools and the liminal experience of their use. *Information Systems Research*, *32*(4), pp.1192-1213.

- Zhao, K. and Xia, M., 2014. Forming Interoperability through Interorganizational Systems Standards. *Journal of Management Information Systems*, *30*(4), pp.269-298.